\documentstyle[12pt,epsf,epsfig]{article}
\newcommand{\summ}[2]{\sum\limits_{#1}^{#2}}
\newcommand{\fitem}[2]{\vspace{0.8cm}{\bf
#1} #2\vspace{0.3cm}}
\newcommand{\nDsum}[2]{\sum_{\stackrel{\scriptstyle #1(\CK{#2})
\in \Z} {\delta #1=0}}}
\newcommand{\corr}[1]{<\!\!#1\!\!>}
\newcommand{\DIM}{D}
\newcommand{\DIMO}{D-1}
\newcommand{\eq}[1]{(\ref{#1})}
\newcommand{\diff}{\partial}
\newcommand{\beq}{\begin{equation}}
\newcommand{\eeq}{\end{equation}}
\newcommand{\beqn}{\begin{eqnarray}}
\newcommand{\eeqn}{\end{eqnarray}}
\newcommand{\dd}{\mbox{d}}
\newcommand{\cR}{{\cal{R}}}

\newcommand{\cZ}{{\cal{Z}}}
\newcommand{\cX}{{\cal{X}}}
\newcommand{\cD}{{\cal{D}}}

\newcommand{\cS}{S}
\newcommand{\dual}{\mbox{}^{\ast}}
\newcommand{\intpi}{\int\limits_{-\pi}^{+\pi}}
\newcommand{\intinf}{\int\limits_{-\infty}^{+\infty}}
\newcommand{\Z}{{Z \!\!\! Z}}
\newcommand{\CK}[1]{\mbox{\scriptsize C}_{\mbox{$\scriptstyle #1$}}}
\newcommand{\dD}{{\cal D}}
\newcommand{\const}{{\rm const.} \cdot}

\newcommand{\nddsum}[2]{\sum_{\stackrel{\scriptstyle \dual
#1(\dual\CK{#2}) \in \Z} {\delta \dual #1=0}}}
\newcommand{\nsum}[2]{\sum_{\scriptstyle #1(\CK{#2})
\in \Z}}
\newcommand{\ndsum}[2]{\sum_{\scriptstyle \dual #1(\dual\CK{#2})
\in \Z}}

\def\NP{ Nucl.~Phys.}

\def\PL{ Phys.~Lett.}

\def\PRp{ Phys.~Rep.}
\def\PR{ Phys.~Rev.}

\newcommand{\abstracts}[1]{{
\centering{\begin{minipage}{12.2truecm}
\normalsize\baselineskip=15pt
\centerline{\footnotesize ABSTRACT}\vspace*{0.3cm}
\parindent=20pt #1
\end{minipage}}\par}}
\begin{document}
~\vspace{-1.5cm}
\begin{flushright}
{\large ITEP-TH-55/97}
\end{flushright}
\vspace{1.5cm}

\begin{center}

{\baselineskip=16pt
{\Large \bf ABELIAN PROJECTIONS AND MONOPOLES}\\

\vspace{1cm}

{\large
M.N.~Chernodub and M.I.~Polikarpov\footnote{Lectures given by
Mikhail Polikarpov at the Workshop "Confinement, Duality and 
Non-Perturbative Aspects of QCD", Cambridge (UK), 24 June - 4 July 1997.}
}\\

\vspace{.5cm}
{ \it
Institute of Theoretical and Experimental Physics, \\
B.~Cheremushkinskaya~25, Moscow, 117259, Russia
}
}
\end{center}

\vspace{1cm}

\abstracts{
The monopole confinement mechanism in the abelian projection of
lattice gluodynamics is reviewed. The main topics are: the abelian
projection on the lattice and in the continuum, a numerical study of
the abelian monopoles in the lattice gauge theory. Additionally, we
briefly review the notation of differential forms,  duality,  and
the BKT transformation in the lattice gauge theories.
}

\newpage

\setcounter{footnote}{0}
\renewcommand{\thefootnote}{\alph{footnote}}

\section{Introduction}

In these lectures we give an introduction to the theory of the
confinement of color in lattice gauge theories. For the sake of
self-consistency, we explain all definitions.  The reader is supposed
to be familiar with  basic notions of lattice gauge theory.

Lattice field theories were originally formulated \cite{Wilson}
in order to explain the confinement of color in nonabelian gauge
theories. The leading  term of the strong coupling ($1\slash g^2$)
expansion in lattice gauge theories yields  the area law for the
Wilson loop. The $D$ dimensional theory is reduced to the
2--dimensional one, the field strength tensors are independent on
different plaquettes and the confinement has a stochastic nature. The
expectation value of the Wilson loop is simply ${<P>}^S$, where $<P>$
is the expectation value of the plaquette, $S$ is the area of the
minimal surface spanned on the loop. Therefore,  the string tension
is $\sigma = - \ln <P>$.

The numerical study of lattice gauge theories initiated by
Creutz \cite{Creutz} shows that the strong coupling expansion of
lattice gauge theory has no relation to continuum physics: the
effects of lattice regularization are very strong, the results
are not rotationally invariant,  and there is no scaling for
hadron masses\footnote{Here, scaling   means  independence of
the  ratios of the hadron masses $m_k \slash m_i$  on  the parameters of the
theory such as  the bare coupling and  the cutoff parameter.}.

\begin{figure}[htb]
\vskip0mm
\begin{center}
\begin{tabular}{cc}
{\epsfxsize=0.45\textwidth\epsfbox{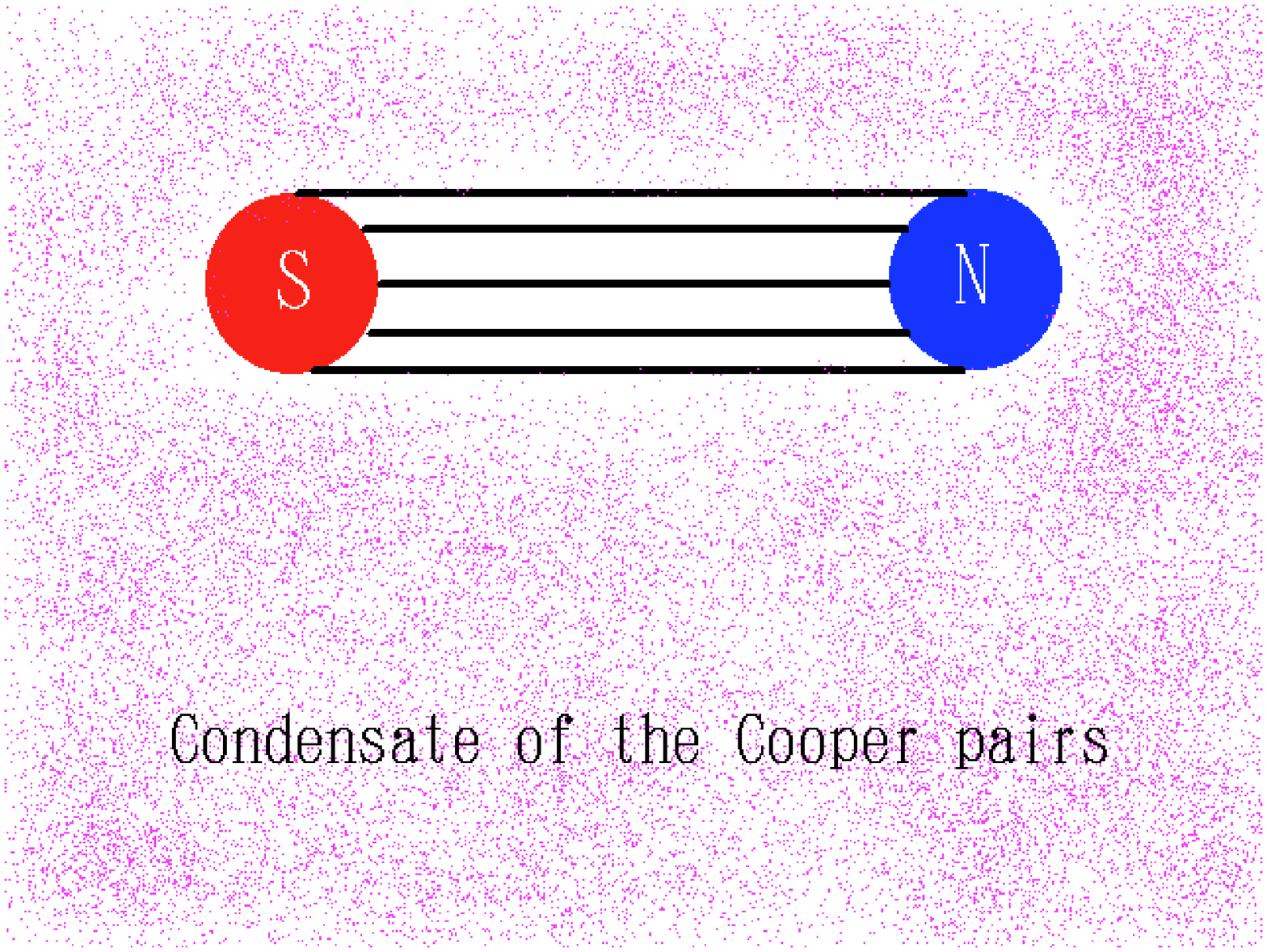}}&
{\epsfxsize=0.45\textwidth\epsfbox{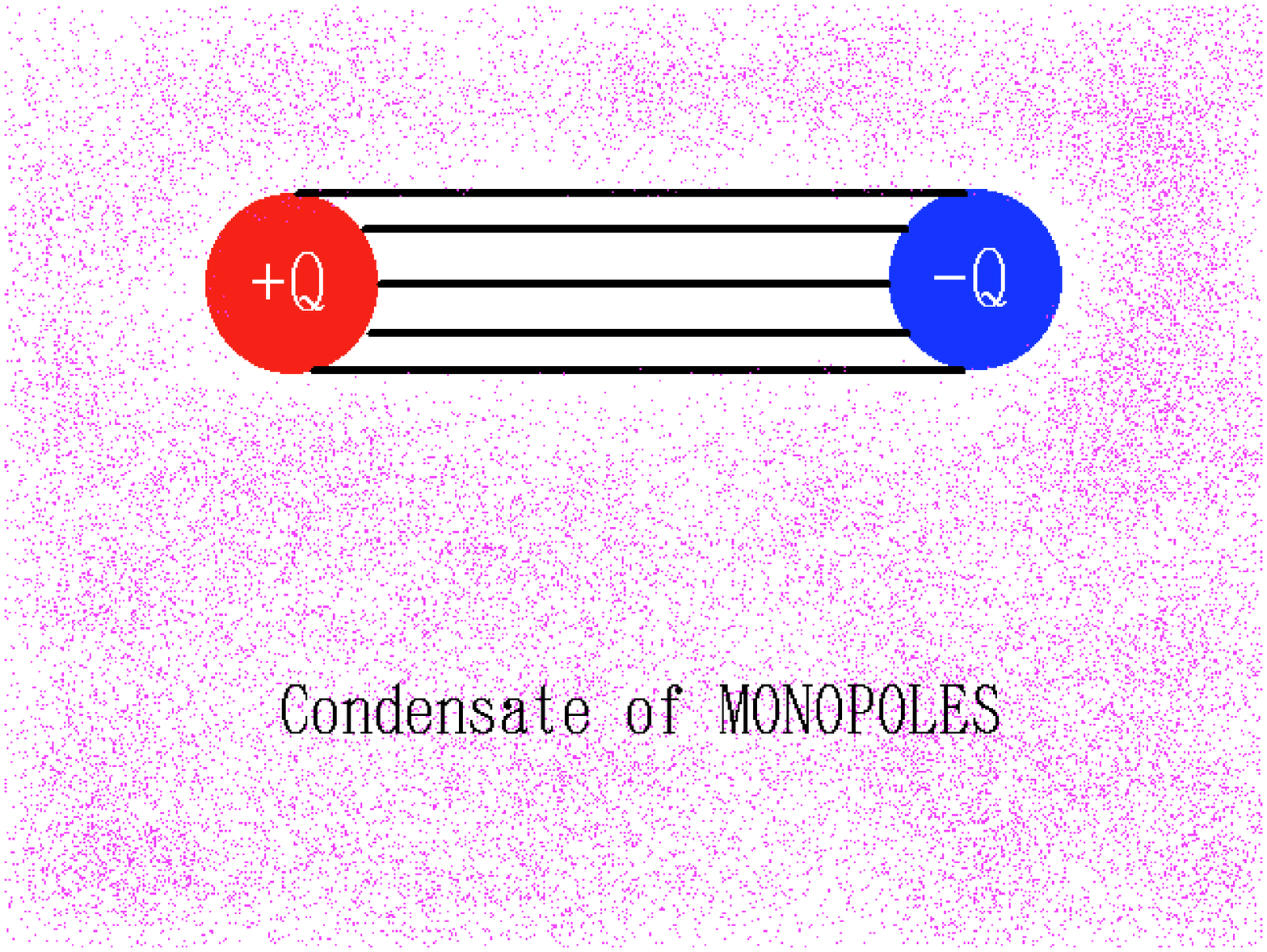}}\\
 (a) & (b)\\
\end{tabular}
\end{center}
\vskip-5mm
\caption{The Abrikosov string between the monopoles in the
superconductor (a) and an  analogue of the Abrikosov string between
the electrically charged particles in the dual superconductor (b).}
\label{superc}
\end{figure}

In the weak coupling region (where the existence of the continuum
limit was found in  numerical calculations),  the confinement
mechanism has no stochastic origin. There are several approaches to
explain color confinement. We will describe  the most traditional
one, namely,  confinement  caused by the dual Meissner effect. The
linear confining potential can be explained by the formation of a
string (flux tube) connecting a quark and an anti-quark. A
well-known example of the string--like solution of the classical
equations of motion is given in Ref. \cite{Abr57}. If we have a
medium of condensed electric charges (a superconductor), then between
the monopole and anti-monopole  an Abrikosov string is formed, see
Figure~\ref{superc}(a). To explain the confinement of electric
charges,  we need a condensate of magnetic monopoles,
Figure~\ref{superc}(b). 
This simple qualitative idea was suggested by
't~Hooft and Mandelstam \cite{MatH76}.  There is a long way from this
picture to real QCD. First,  we have to explain how the abelian gauge
field with monopoles is obtained  from the non-abelian gauge field.
Secondly,  we have to explain why   a  dual superconductor is
involved here. We discuss these two questions in these lectures.

At  present,  we have no analytic proof of the existence of
the condensate of abelian magnetic monopoles in gluodynamics
and in chromodynamics. However, in all  theories allowing for an
analytical proof of confinement, the latter  is due
to the condensation of monopoles. These analytical examples are:
compact electrodynamics \cite{PolyakovQED}, the Georgi--Glashow
model \cite{PolyakovGG},  and super-symmetric Yang--Mills
theory \cite{SeibergWitten}.

On the other hand,  many numerical facts (some of these are
discussed  in these lectures) suggest that the vacuum in $SU(2)$ and $SU(3)$
lattice gauge theories behaves as  a dual superconductor. As an
illustration,  we give  two figures obtained by numerical calculations
in $SU(2)$ lattice gluodynamics.

\begin{figure}[t!]
\vskip5mm
\centerline{\epsfxsize=0.7\textwidth\epsfbox{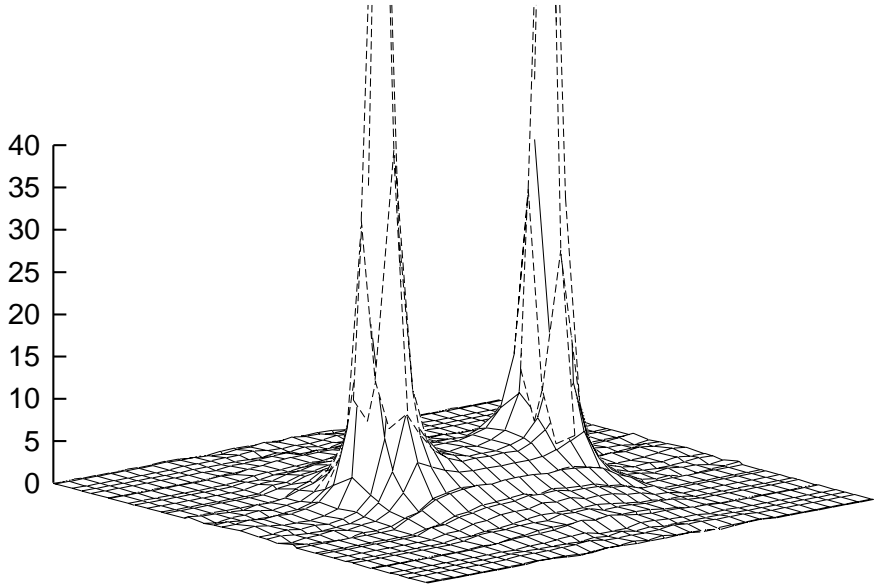}}
\vskip3mm
\caption{The
action density distribution \cite{bali0} at $\beta=2.635$, the
distance between quark and anti-quark sources is $\approx 1.35 fm$.
Two horizontal axes correspond to two spatial lattice axes.}
\label{bali1}
\end{figure}

\begin{figure}[b!]
\vskip1cm
\centerline{\epsfxsize=0.6\textwidth\epsfbox{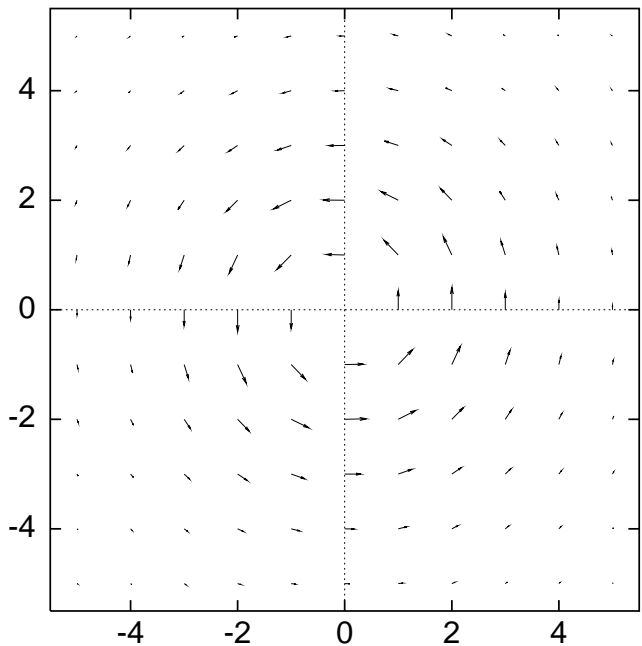}}
\vskip-7mm
\caption{The monopole currents around the string tube which is formed
between the static quark and anti-quark in gluodynamics,
Ref. \cite{bali2}.}
\label{bali2}
\end{figure}

In Figure~\ref{bali1},  taken from Ref. \cite{bali0}, the action
density (vertical axis) of the $SU(2)$ fields is shown. The two peaks
correspond to the quark--anti-quark pair, the formation of the flux
tube is clearly seen.  In Figure~\ref{bali2}, taken from
Ref. \cite{bali2}, the abelian monopole currents near the center
of the flux tube formed by the  quark--anti-quark pair are shown. It
is seen that the monopoles  wind around the center of the flux tube
just as the Cooper pairs wind around the center of the Abrikosov
string.

We first explain how to get the abelian fields and monopoles from the
non-abelian fields. Then we present the results of numerical studies of
the confinement mechanism in lattice gluodynamics.

All technical details such as the  formalism of  differential forms on the
lattice are given in the Appendices.

\section{Abelian Monopoles from Non-Abelian Gauge Fields}

In this Section we discuss the question how  to obtain the abelian
monopoles from non-abelian gauge fields.

\subsection{The Method of Abelian Projection }

The abelian monopoles arise from non-abelian gauge fields
as a result of
the abelian projection suggested by
`t~Hooft \cite{tH81}. The abelian
projection is a partial gauge fixing under which the abelian degrees
of freedom remain  unfixed. For example, the abelian projection of a
theory with $SU(N)$ gauge symmetry leads to a theory with
${[U(1)]}^{N-1}$ gauge symmetry.

Since the original $SU(N)$ gauge symmetry group is compact, the
remaining abelian gauge group is also compact. But the abelian gauge
theories with compact gauge symmetry group possess  abelian
monopoles. Therefore $SU(N)$ gauge theory in the abelian gauge
has  abelian monopoles.

First,  consider the   simplest example of {\it  $F_{12}$ abelian
projection} for $SU(2)$ gauge theory. This gauge is defined by
the following condition:
\beqn
  \hat{F}_{12}(x) = {\rm diagonal \, matrix}\,. \label{F12}
\eeqn
It is easy to fix to this gauge, since under gauge transformations
the field strength tensor $\hat{F}_{12}$ transforms as
\beqn
 \hat{F}_{12}(x) \to \hat{F}_{12}'(x) = \Omega^+(x) \hat{F}_{12}(x)
 \Omega (x)\,.
\eeqn
If we fix to the $F_{12}$ gauge, then the field strength tensor
$\hat{F}_{12}'(x)$ is invariant under $U(1)$ gauge
transformations:
\beqn
 {\hat F}'_{12}(x) = \Omega^+_{U(1)}(x)  \hat{F}'_{12}(x)
 \Omega_{U(1)}(x)\,,
\eeqn
where
\beqn
\Omega_{U(1)} (x) = \pmatrix{
e^{i\alpha(x)} & 0 \cr 0 & e^{-i\alpha(x)} \cr}\,, \quad
\alpha\in[0,2\pi)\,.
\label{U1}
\eeqn
Therefore, the gauge condition \eq{F12} fixes the $SU(2)$ gauge group
up to the diagonal $U(1)$ subgroup.

The $SU(2)$ gauge field
\beqn
\hat{A}_\mu = \pmatrix{ A^3_\mu & A_\mu^+ \cr A_\mu^- & -A^3_\mu \cr}
\eeqn
transforms under the gauge transformations as
\beqn
\hat{A}_\mu \rightarrow \Omega^+ \hat{A}_\mu \Omega -\frac ig
\Omega^+\partial_\mu \Omega\,. \label{Ar}
\eeqn

If we fix to the $F_{12}$ gauge, then
under the remaining $U(1)$ gauge transformation \eq{U1} the
components of the nonabelian gauge field $\hat{A}$ transform as
\beqn
a_\mu \rightarrow  a_\mu - \frac 1g \partial_\mu \alpha\,,
\quad a_\mu = A_\mu^3\,;\qquad
A_\mu^+  \rightarrow  e^{2i\alpha} A_\mu^+\,,\quad
A^+_\mu = A^1_\mu + i A^2_\mu\,.
\label{U1c}
\eeqn
Thus,  in the abelian gauge  the field $a_\mu$ plays the role of the
abelian gauge field and the field $A_\mu^+$ is a charge 2 abelian
vector matter field.

We thus obtain abelian fields from non-abelian ones. It occurs
that  abelian projection is also responsible
for the appearance of abelian monopoles.  If
$a$ is a  regular abelian field, then $\mbox{div}\vec{H} = 0$, since
$\vec{H} = \mbox{curl}\,\vec{a}$. But the abelian gauge field may be
non-regular, since the matrix of the gauge transformation $\Omega$
may contain singularities. The nonabelian field strength tensor
$\hat{F}_{\mu\nu}$ is {\it not invariant} under  singular gauge
transformations:
\beqn
\hat{F}_{\mu\nu}[A] \to \hat{F}_{\mu\nu}[A^{(\Omega)}] =
\Omega^+ \hat{F}_{\mu\nu}[A] \Omega +
\hat{F}^{sing}_{\mu\nu}[\Omega]\,,
\label{8a}
\eeqn
where
\beqn
\hat{F}^{sing}_{\mu\nu}[\Omega]  =
- \frac ig \Omega^+(x) [ \partial_{\mu} \partial_{\nu} -
\partial_{\nu} \partial_{\mu} ] \Omega(x)\,.
\eeqn

Therefore,  if we fix to the abelian gauge~\eq{F12},  using the singular
gauge rotation matrices,  the {\it abelian} field strength tensor
\beqn
f_{\mu\nu} = \partial_\mu a_\nu - \partial_\nu a_\mu
\eeqn
may contain singularities (the Dirac strings)
\beqn
f_{\mu\nu} = f^r_{\mu\nu} + f^s_{\mu\nu}\,, \qquad
\varepsilon_{\mu\nu\alpha\beta} \diff_\nu f^r_{\alpha\beta} = 0\,
\quad {\rm and}
\quad \varepsilon_{\mu\nu\alpha\beta} \diff_\nu f^s_{\alpha\beta}
\neq 0\,,
\eeqn
and therefore,  $\mbox{div}\vec{H} \neq 0$.

The charge of the monopole can be calculated \cite{KrScWi87}
by means  of the Gauss law.  Let us choose an abelian monopole in a
certain  time slice and surround it by an infinitesimally small
sphere $S$.  The monopole charge is
\beqn
m & = & \frac{1}{4\pi}\oint \vec{H} \, d \vec{\sigma} =
\frac{1}{8 \pi}{\int} d\sigma_{\mu\nu}\,
\varepsilon_{\mu\nu\alpha\beta} f_{\alpha\beta}
= \frac{i}{4 \pi g}{\int} d\sigma_{\mu\nu}\,
\varepsilon_{\mu\nu\alpha\beta} {\rm Tr} [\Omega^+\diff_{\alpha}
\Omega\Omega^+\diff_{\beta}\Omega] \nonumber\\
&=&-\frac{i}{4 \pi g}{\int} d\sigma_{\mu\nu}\,\varepsilon_{\mu\nu\alpha\beta}
\diff_\alpha {\rm Tr} [\Omega^+\diff_{\beta}\Omega]
= 0,\pm \frac{1}{2 g}, \pm \frac{1}{g},\dots\,.
\label{mcharge}
\eeqn
The quantization of the abelian monopole charge is due to
topological reasons: the surface integral in eq.\eq{mcharge} is equal to
the winding number of $SU(2)$ over the sphere $S$ surrounding the monopole.
The physics of quantization is simple: the electric charge is fixed
by the gauge transformation~\eq{U1c}, and magnetic charge should obey
the Dirac quantization condition. The last integral can be seen as
the magnetic flux through the Dirac string that arises as a consequence
of the gauge singularity.

\subsection{Various Abelian Projections}

There is  an infinite number of abelian projections. In the
previous section we  have considered  the $\hat{F}_{12}$ abelian gauge.
Instead of the diagonalization of the tensor component $\hat{F}_{12}$
by the gauge transformation,  we can diagonalize any operator $X$ which
transforms under the gauge rotation as follows:  $X \to \Omega^+ X
\Omega$. Each operator $X$ defines an abelian projection. At finite
temperature one can consider the so-called {\it Polyakov abelian
gauge} which is defined by the diagonalization of the Polyakov line.

The   most interesting numerical results are those  obtained in the Maximal
Abelian (MaA) gauge. This gauge is defined by the maximization of the
functional
\beqn
  \max_\Omega R[\hat{A}^\Omega]\,,\qquad
  R[\hat{A}] = - \int\, d^4 x \,
  [(A_\mu^1)^2 + (A_\mu^2)^2]\,,
\eeqn

The  condition of  a local extremum   of the functional $R$ is
\beqn
(\diff_\mu \pm i g A^3_\mu) A^\pm =0.
\eeqn
Clearly,  this condition (as well as the functional $R[A]$)
is invariant under the $U(1)$ gauge transformations \eq{U1c}. The
meaning of the MaA gauge is simple: by gauge transformations we  make
the field ${\hat A}_\mu$ as diagonal as possible.

\subsection{Abelian Projection on the Lattice}

The $SU(2)$ gauge fields $U_l$ on the lattice are defined by $SU(2)$
matrices  attached to the links $l$. These lattice fields are
related to the continuum $SU(2)$ fields $\hat A$: $U_{x,\mu} =
e^{i a g \hat{A}_\mu (x)}$, here $a$ is the lattice spacing. Under
the gauge transformation,  the field $U_l$ transforms as
$U_{x,\mu}^\Omega = \Omega_x^+ \, U_{x,\mu} \, \Omega_{x+\hat\mu}$,
the matrices of the gauge transformation are attached to the sites
$x$ of the lattice.

The Maximal Abelian gauge is defined on the lattice by the following
condition \cite{KrScWi87}:
\beqn
\max_\Omega R[\hat{U}^\Omega_l]\,,\qquad
R[U_l] = \sum\limits_l Tr[\sigma_3 U_l^+ \sigma_3 U_l]\,,
\quad l = \{x,\mu\}\,.
\eeqn
This gauge condition corresponds to an abelian gauge,  since $R$ is
invariant under the gauge transformations defined by the matrices~\eq{U1}.

Let us parametrize the link matrix $U$ in the standard way
\beqn
U_l = \pmatrix{
\cos \varphi_l e^{i\theta_l} & \sin \varphi_l e^{i \chi_l}\cr
- \sin \varphi_l e^{- i \chi_l} & \cos \varphi_l
e^{-i\theta_l} \cr}\,, \label{LatticeU}
\eeqn
where $\theta,\chi\in[-\pi,+\pi)$ and $\varphi\in[0,\pi)$. In this
parameterization,
\beqn
R[U_l] = \sum_l \cos 2 \varphi_l\,.
\eeqn
Thus,  the maximization of $R$ corresponds to the maximization of
the diagonal elements of the link matrix~\eq{LatticeU}.

Under the $U(1)$ gauge transformations,  the components of the gauge
field \eq{LatticeU} are transformed as
\beqn
\theta_{x,\mu} \rightarrow \theta_{x,\mu}
+  \alpha_x - \alpha_{x+\hat\mu}\,,\quad
\chi_{x,\mu} \rightarrow \chi_{x,\mu}
+  \alpha_x + \alpha_{x+\hat\mu}\,,\quad
\varphi_{x,\mu} \rightarrow \varphi_{x,\mu}
\eeqn
Therefore,  in the MaA the gauge,  the field $\theta$ is the $U(1)$ gauge
field, the field $\chi$ is the abelian charge 2 vector matter
field, the field $\varphi$ is the non--charged vector matter field.

\subsection{Monopoles on the Lattice}

\begin{figure}[htb]
\vskip5mm
\begin{center}
\begin{tabular}{cc}
{\epsfxsize=0.48\textwidth\epsfbox{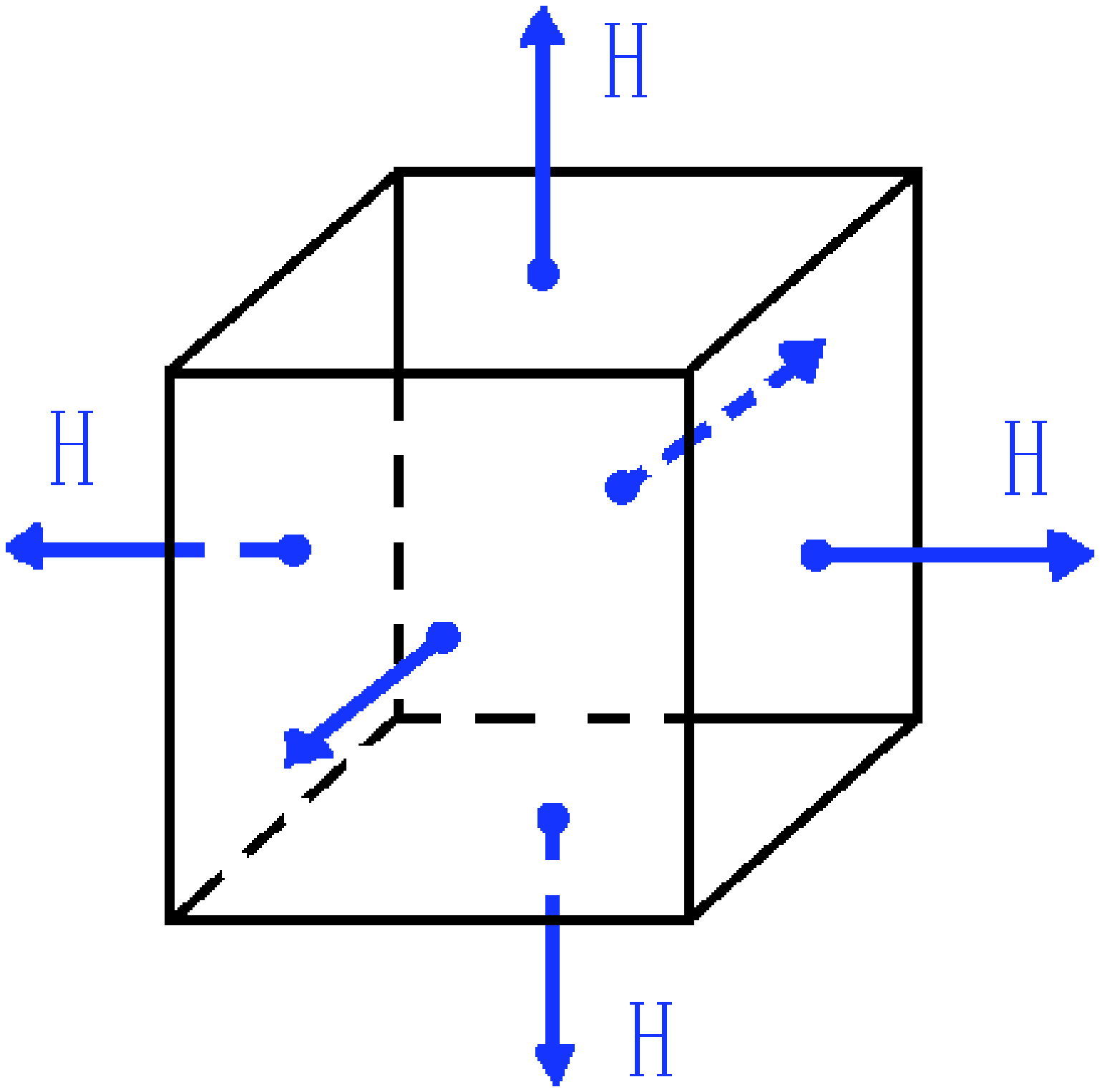}}& \vspace{.2cm}
{\epsfxsize=0.3\textwidth\epsfbox{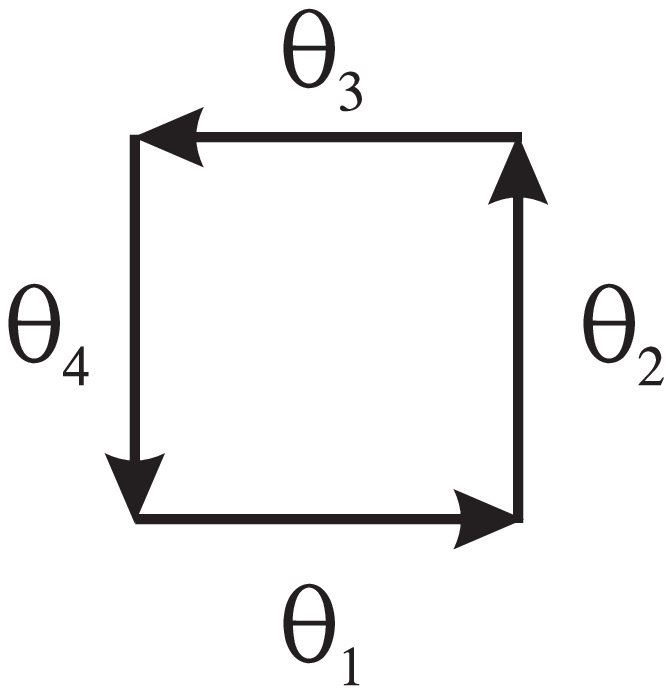}}\\
 (a) & (b)\\
\end{tabular}
\end{center}
\vskip-3mm
\caption{The magnetic flux through the boundary of the
cube $C$ (a) and lattice field strength tensor (b).}
\label{cube0}
\end{figure}

A  configuration of  abelian gauge fields $\theta_l$ can
contain monopoles. The position of the monopoles is defined by the
lattice analogue of the Gauss theorem. Consider the elementary
three--dimensional cube $C$ (Figure~\ref{cube0}(a)) on the lattice.

The abelian magnetic flux $\vec{H}$ through the surface of
the cube $C$ is given by the   formula
\beqn
 m = \frac{1}{2 \pi} \sum\limits_{P\in\partial C}
 {\bar \theta}_P\,,
 \label{LatticeFlux}
\eeqn
where ${\bar \theta}_P$ is the magnetic field defined as follows.
Consider the plaquette angle $\theta_P  = \theta_1 + \theta_2 -
\theta_3 - \theta_4 \equiv \dd \theta$, the $\theta_i$'s are attached
to the links $i$ which form the boundary of the plaquette $P$,
Figure~\ref{cube0}(b). The definition of ${\bar \theta}_P$ is ${\bar
\theta}_P = \theta_P + 2 \pi k$, where the integer $k$ is such that $-
\pi < {\bar \theta}_P \le \pi$. The restriction of ${\bar \theta}_P$
to the interval $(-\pi,\pi]$ is natural since (as we point out in
Appendix~A)  the abelian action for the compact fields $\theta_l$ is a
periodic function of ${\bar \theta}_P$. Equation~\eq{LatticeFlux} is
the lattice analogue of the continuum formula  $m = \oint \vec{H} \,
{\rm d} \vec{S}$. Due to the compactness of the lattice field
$\theta$ ($-\pi < \theta \le \pi$) there exist singularities (Dirac
strings), and therefore,  $\mbox{div}\vec{H} \neq 0$.
\vskip3mm

\begin{figure}[htb]
\centerline{\epsfxsize=0.48\textwidth\epsfbox{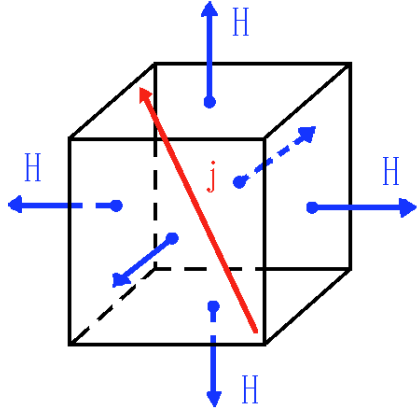}}
\vskip-3mm
\caption{The definition of the magnetic monopole current.}
\label{monopolecurrent}
\end{figure}

The magnetic charge $m$ defined by eq.~\eq{LatticeFlux} has the
following properties:
\begin{enumerate}
\item $m$ is quantized: $m = 0,\pm 1,\pm 2$;
\item If $m \neq 0$,  then there exists a magnetic current $j$. This
current is attached to the link  dual to cube $C$,
Figure~\ref{monopolecurrent}.
\item Monopole currents $j$ are conserved:  $\delta j = 0$, the
currents form closed loops on the $4D$ lattice. The proof is given in
Example~3 of Appendix~A.
\end{enumerate}

\subsection{Vacuum of Gluodynamics As a (Dual) Superconductor}

As we  have just shown,  the non-abelian gauge field in  the abelian
projection is reduced to abelian fields and abelian monopoles. The
static quark--anti-quark pair in   the abelian projection becomes the
electric charge--anti-charge pair, and  if  the monopoles are condensed,
then the quark and anti-quark are connected to each other by an
analogue of the Abrikosov string. A  more detailed discussion of the
confinement in the abelian projection is given in
Refs. \cite{tH81}.  Thus,  in order to justify  the monopole
confinement mechanism we have to prove the existence of the monopole
condensate.  For lattice gluodynamics we have a lot of numerical facts
which confirm  the monopole confinement mechanism. We discuss these
 at the end of this section. But  first we give an analytical
example which shows how the monopole condensate appears in compact
electrodynamics.

\subsubsection{Compact $U(1)$ Lattice Gauge Theory}

We show that lattice compact electrodynamics can be represented
as a dual Abelian Higgs model,  with the Higgs particles being  monopoles. The
partition function of the compact $U(1)$ gauge theory can be written
as
\beqn
\cZ = \int\limits_{-\pi}^{+\pi} \cD
\theta_l \, \exp\{ - \cS( \dd \theta)\}\,,
\label{PartFunU1}
\eeqn
where,  as below,  $\int\limits_{-\pi}^{+\pi} \cD
\theta_l = \prod\limits_l \int\limits_{-\pi}^{+\pi} \dd \theta_l$ is
the integral over all link variables.
In the continuum limit ($a \to 0$),
$\cS(\theta_P) \propto f^{\,2}_{\mu\nu}\, a^4$.

\begin{figure}[htb]
\vskip3mm
\centerline{\epsfxsize=0.30\textwidth\epsfbox{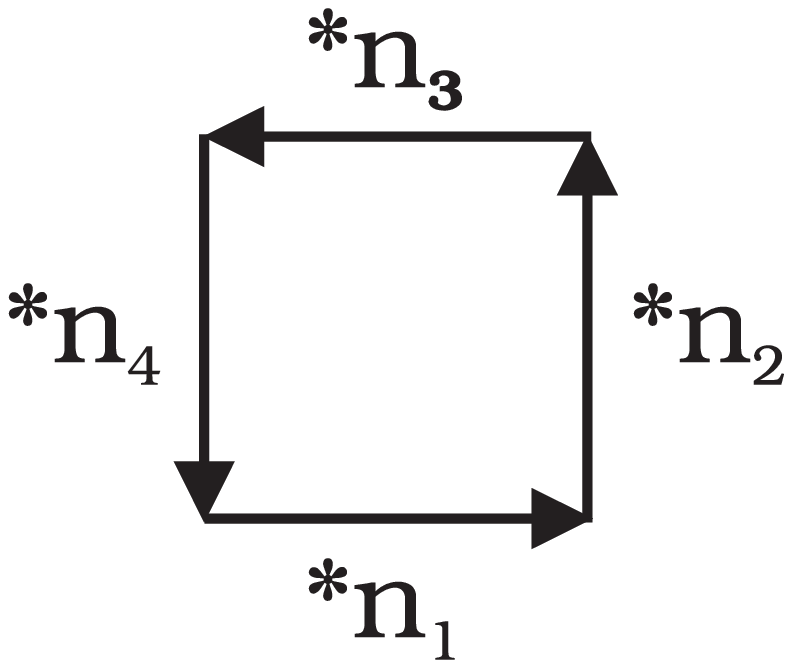}}
\vskip-3mm
\caption{Field strength tensor on the dual lattice.}
\label{dualFST}
\end{figure}

By means of the duality transformation (see Appendix~B), the theory
\eq{PartFunU1} can be rewritten in the form
\beqn
\cZ =\const\nddsum{n}{1} \, \exp\{ - \cS^d (\dd \dual n)\}\,,
\label{U1PFDual}
\eeqn
where $\dd \dual n = \dual n_1 + \dual n_2 - \dual n_3 - \dual n_4$
is the plaquette constructed from the integers $n$  (see
Figure~\ref{dualFST}). The dual action is defined by eq.\eq{FT}.

The integer valued variable $n_{x,\mu}$ is dual to the field
$\theta_{x,\mu}$. In the continuum limit,  $\theta_{x,\mu} \propto
A_\mu (x)$ is the ordinary $U(1)$ gauge field and $n_{x,\mu}\propto
B_\mu(x)$ is the dual $U(1)$ gauge field\footnote{In the continuum
notations we do not use the symbol ``$\dual$" to denote the dual
fields.}, $\partial_\mu A_\nu - \partial_\nu A_\mu = \frac{1}{2}
\varepsilon_{\mu\nu\alpha\beta} \partial_\alpha B_\beta$.

The partition function \eq{U1PFDual} can be represented as that
 of the (dual) Abelian Higgs model.
\beqn
\cZ =\const\lim\limits_{\gamma \to \infty}
\lim\limits_{\lambda \to \infty}
\int\limits^{+\infty}_{-\infty} \cD \dual B \int\limits
\cD \dual \Phi\, \exp\{ - \cS^{AH}(\dual B, \dual \Phi)\}\,,
\label{PAH}
\eeqn
where the action of the (dual) Abelian Higgs model is
\beqn
\cS^{AH}(\dual B, \dual \Phi) =
\sum\limits_P S^d (\dd \dual B) + \frac{1}{2}
\sum\limits_x \sum^4_{\mu=1} {|\dual \Phi_x - e^{i \dual B_{x,\mu}}
\dual \Phi_{x+\hat\mu}|}^2 + \lambda \sum
(|\dual \Phi_x|^2 - \gamma)^2\,.
\label{SAH}
\eeqn
Here $\dd \dual B$ is the plaquette variable constructed from the
link variable $\dual B_l$, where $\dual B_l$ is the dual gauge field.
In the continuum limit,  we have $S(\dd \dual B) \to \beta {(\partial_\mu B_\nu -
\partial_\nu B_\mu)}^2$. The second term in eq.\eq{SAH} has the
following continuum limit: ${|\Phi_x - e^{i \dual B_{x,\mu}} \dual
\Phi_{x+\hat\mu}|}^2 \to {|(\partial_{\mu} - i B_{\mu}) \Phi|}^2$. The
action $S^{AH}$ is invariant under the following gauge
transformations: $\dual B \to \dual B - \dd \dual \alpha$, $\dual
\Phi \to \dual \Phi e^{i \dual \alpha}$.

In the London limit,  $\lambda \to \infty$,   the mass of the Higgs
particle tends to infinity. The radius of the Higgs field $|\dual
\Phi|$ is fixed and only the phase $\dual \varphi$ of the Higgs field
$\dual \Phi= |\dual \Phi| e^{i\dual \varphi}$ is a physical degree of
freedom. In the unitary gauge $\dual \varphi=0$, the action of the
Abelian Higgs model \eq{SAH} is
\beqn
\cS^{AH} = \sum\limits_P S^d (\dd \dual B) + \frac{\gamma}{2}
\sum\limits_x \sum\limits^4_{\mu=4} {|1 - e^{i \dual
B_{x,\mu}}|}^2\,.
\label{pf2}
\eeqn

In the limit $\gamma \to \infty$,  the photon mass
$m_{ph}^2=\gamma/\beta$ becomes infinite and the field $B_{x,\mu}$ is
equal to zero modulo $2 \pi$: $\dual B_{x,\mu} = 2 \pi
\dual n_{x,\mu}$. In this limit,  the partition function \eq{pf2}
reduces to
\beqn
\cZ =\const\nddsum{n}{1} e^{- S^d ({\rm d} \dual n)}\,,
\qquad S(\dd \dual n) = 4\pi^2\, \beta \, {(\dd \dual n)}^2\,.
\eeqn

Therefore,  the compact $U(1)$ gauge theory is equivalent to the dual
abelian Higgs model in the double limit
\beqn
{\rm Compact \,} U(1) {\, \rm gauge \, theory} \, = \,
\lim\limits_{\gamma \to \infty}
\lim\limits_{\lambda \to \infty} {\rm (Abelian \, Higgs \, model)}
\eeqn

The gauge fields $B_l$ in eq.\eq{SAH} are dual to the original
gauge fields $\theta_l$, and these  interact via the covariant
derivative with the Higgs field $\Phi$. Therefore,  the field $\Phi$
carries  the magnetic charge, and due to the Higgs potential in
eq.\eq{SAH},  these monopoles are condensed at the classical level.
It is well known that in the quantum $4D$ compact electrodynamics
there exists a confinement--deconfinement phase transition. It
can be shown by numerical calculations \cite{PoWi,PoPoWi91} that in
the confinement phase the monopoles are condensed and that in the
deconfinement phase the monopoles are not condensed.

\subsubsection{What is the Theory Dual to Gluodynamics?}

In any abelian projection, lattice gluodynamics corresponds to
some abelian gauge theory. This abelian theory, in general,  is very
complicated and  non-local. Nevertheless,   in the Maximal
Abelian projection,   numerical calculations show that the abelian
monopoles are important degrees of freedom, and that they are responsible for
the confinement. Moreover, as we show in the next section (see also
Refs. \cite{TS,SuzEtAll}),  the distribution of
monopole currents indicates that, at large distances, gluodynamics is
equivalent to the dual Abelian Higgs model, the Higgs particles are
abelian monopoles and these  are condensed in the confinement phase.

\section{Numerical Facts in $4D$ $SU(2)$ Gluodynamics}

The standard scheme of  numerical calculations can be described  as follows.

\begin{figure}[htb]
\vskip1.7cm
\begin{center}
\begin{tabular}{cc}
\hskip-3mm{\epsfxsize=0.4\textwidth\epsfbox{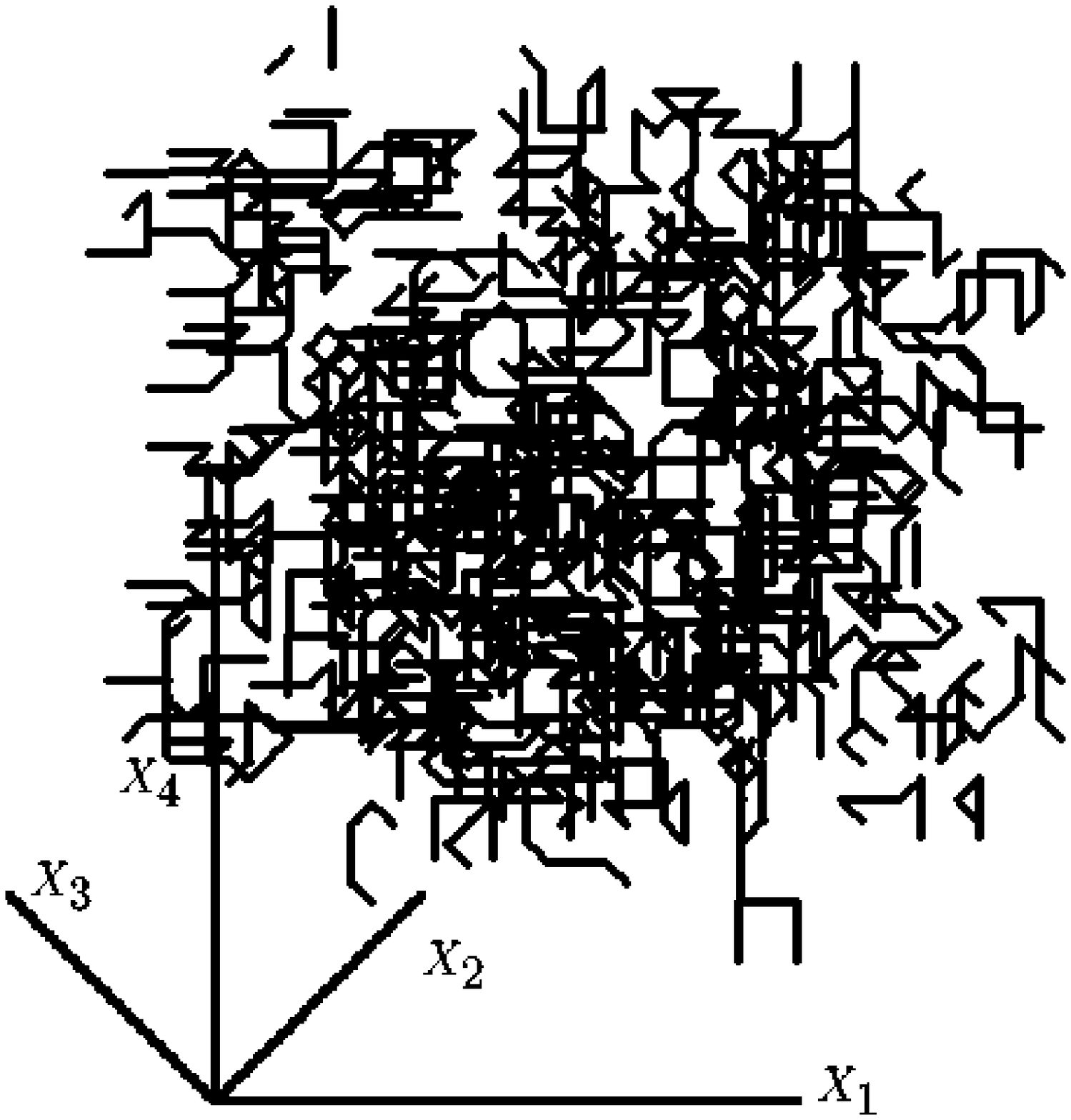}}&\hskip8mm
{\epsfxsize=0.4\textwidth\epsfbox{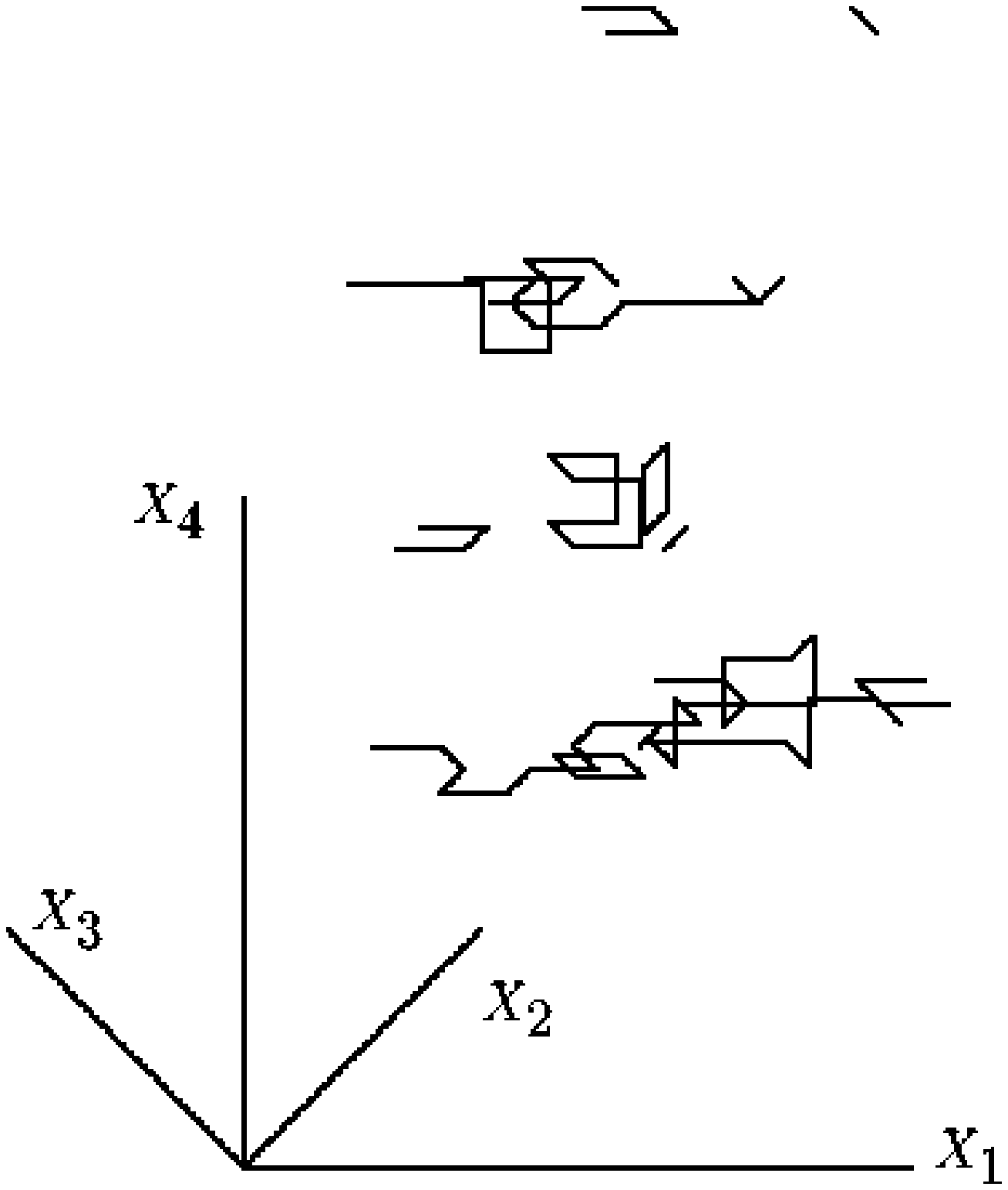}}\\
 (a) & (b)\\
\end{tabular}
\end{center}
\vskip-5mm
\caption{The abelian monopole currents for the confinement (a)
($\beta=2.4$, $10^4$ lattice) and the deconfinement (b) phases
($\beta=2.8$, $12^3 \cdot 4$ lattice).}
\label{CurrentsConf}
\end{figure}

\begin{enumerate}
\item Generate the lattice Yang--Mills fields $U_l$ using the
standard Monte-Carlo method. Thus we obtain the configurations of
fields distributed according to the Boltzmann factor
$e^{-S^{YM}}$;

\item Perform the abelian gauge fixing and extract abelian gauge
fields from the non-abelian ones. In the case of the MaA projection,
the abelian gauge fixing is a rather time-consuming problem.

\item Extract abelian monopole currents from the abelian gauge
fields. As  mentioned above,  the monopole currents form
closed paths on the dual lattice. In Figure~\ref{CurrentsConf} we
show the abelian monopole currents for the confinement (a) and
the deconfinement (b) phases. It is
seen that in the confinement phase the monopoles form a  dense cluster,
and  there is  a  number of small  mutually disjoint  clusters. In the
deconfinement phase the monopole currents are dilute.

\item Calculate expectation values of various operators using the
monopole currents. Below we discuss a number of numerical facts
which show that abelian monopoles in the MaA gauge are the appropriate
degrees of freedom to describe confinement.

\end{enumerate}

\subsection{Fact~1: Abelian and Monopole Dominance}

The notion of the ``abelian dominance'' introduced in
Ref. \cite{SuYo90} means that the expectation value of a
physical quantity $<\cX>$ in the nonabelian theory coincides with (or
is  very close to) the expectation value of the corresponding abelian
operator in the abelian theory obtained by abelian projection.
Monopole dominance means that the same quantity can be calculated
in terms of the monopole currents extracted from the abelian fields.
If we have $N$ configurations of  nonabelian fields on the lattice,
the abelian and monopole dominance means that \beqn \frac1N\sum_{conf}\cX(\hat
U_{nonabelian}) = \frac1N\sum_{conf}\cX '(U_{abelian})=
\frac1N\sum_{conf}\cX ''(j) \, . \label{abdom}
\eeqn
Here each sum is taken over all configurations;
$U_{abelian} = e^{i\theta_l}$ is the abelian part of the nonabelian
field $\hat U_{nonabelian}$, $j$ is the monopole current extracted
from $U_{abelian}$. It is clear that $\frac1N\sum_{conf}\cX(\hat
U_{nonabelian})$ is a gauge invariant quantity, while the abelian
and the monopole contributions depend on the type of abelian
projection. In numerical calculations the equalities \eq{abdom} can
be satisfied only approximately.

Among the well-studied problems is that  of  the abelian and the
monopole dominance for the string tension
 \cite{SuYo90,HiKiKi91,StNeWe94,BaBoMu96}. In this case, $\cX(\hat
U_{nonabelian}) = \sigma_{SU(2)}$, $\cX(U_{abelian}) $ $ =
\sigma_{U(1)}$ and the string tension $\sigma_{SU(2)}$
($\sigma_{U(1)}$) is calculated by means of the nonabelian (abelian)
Wilson loops, $Tr \prod_{l\in C} \hat U_l$ ( $\prod_{l\in C}
e^{i\theta_l}$). An accurate numerical study of the MA projection of
$SU(2)$ gluodynamics on the $32^4$ lattice at $\beta = 2.5115$ is
performed in Ref. \cite{BaBoMu96}.

\begin{figure}[t!]
\vskip5mm
\centerline{\epsfxsize=0.65\textwidth\epsfbox{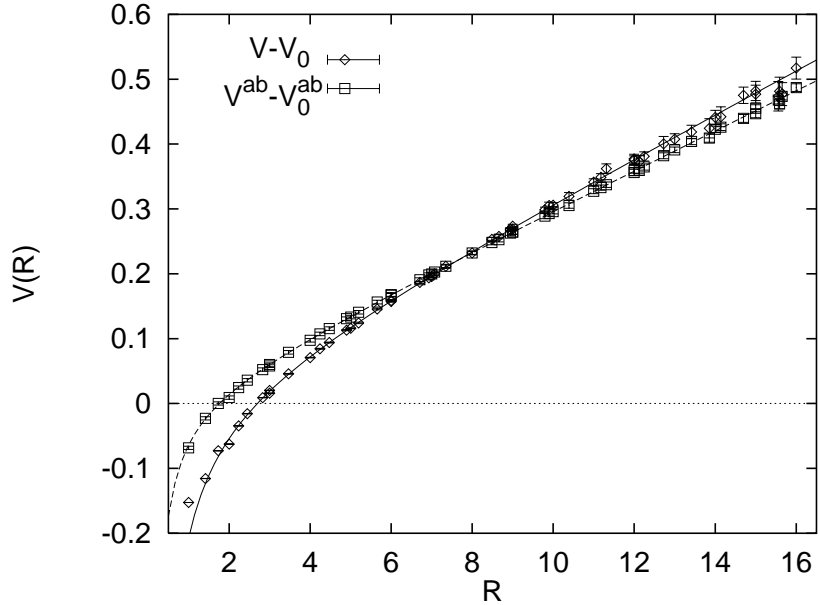}}
\caption{Abelian and nonabelian potentials (with the self
energy $V_0$ subtracted), Ref. \cite{BaBoMu96}.}
\label{bornfull}
\end{figure}

\begin{figure}[t!]
\centerline{\epsfxsize=0.65\textwidth\epsfbox{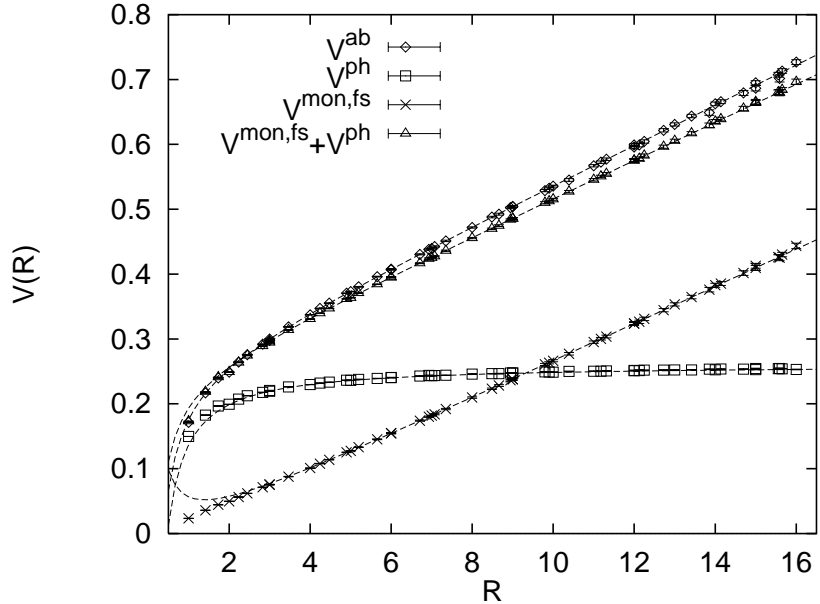}}
\caption{The abelian potential (diamonds) in comparison
with the photon contribution (squares), the monopole contribution
(crosses) and the sum of these two parts (triangles),
Ref. \cite{BaBoMu96}.}
\label{bornfin}
\end{figure}

The corresponding string tension can be taken from the potential
between a heavy quark and antiquark:    $V(r)= V_0 + { \sigma} r - \frac
er$, where $r$ is the distance between the quark and antiquark. The
abelian and the nonabelian potentials are shown in
Figure~\ref{bornfull}. The contribution
of the photon and the monopole parts to the abelian potential is
shown in Figure~\ref{bornfin}.


The differences in the slopes of the linear part of the potentials in
Figure~\ref{bornfull} and Figure~\ref{bornfin} yield the following
relations:  $\sigma_{U(1)} \approx 92\% \, \sigma_{SU(2)}$, $\sigma_j
\approx 95\% \, \sigma_{U(1)}$, where $\sigma_j$ is the monopole
current contribution to the string tension. It is important to study
a widely discussed idea that in the continuum limit ($\beta \to
\infty$),  the abelian and the monopole dominance is exact \eq{abdom}:
$\sigma_{SU(2)} = \sigma_{U(1)} = \sigma_j$.

There are many examples of abelian and monopole dominance in
the MA projection.  The monopole dominance for the string tension has
been found for  the $SU(2)$ positive plaquette model in which $Z_2$
monopoles are suppressed \cite{StNe96}, and also  for the
$SU(2)$ string tension at finite temperature \cite{StWeNe96}, and for
the string tension in $SU(3)$ gluodynamics \cite{Yee94}.
Abelian and monopole dominance for $SU(2)$ gluodynamics has
been found in \cite{Kit96,Eji96} for the Polyakov line and for
critical exponents for the Polyakov line, for the value of the quark
condensate, for the topological susceptibility and also for the hadron
masses in quenched $SU(3)$ QCD with Wilson fermions \cite{Suzm96}.

\subsection{Fact~2: London Equation for Monopole Currents}

In the ordinary superconductor the current of the Cooper pairs
satisfies the London equation. If the vacuum of gluodynamics behaves
as  a dual superconductor,  then it is natural to assume that the abelian
monopole currents should satisfy the dual London equation in the
presence of the dual string
\beqn
\vec{E} - \delta^2\, \mbox{curl}\, \vec{j}_m = \Phi_m \delta (x_x)
\delta (x_y) {\vec n}_z\,,
\eeqn
where $\vec{E}$ is the abelian electric field, $\delta^{-1} = m_{ph}$
is the dual photon mass, $\Phi_m=\frac{2 \pi}{g_m}$ is the electric
flux of the dual string and $g_m$ is the magnetic charge of the
monopole. The string is placed along  the $oz$ axis, the vector ${\vec
n}_z$ is parallel to the direction of the string and we assume
for simplicity that the core of the string is a delta--function.

Indeed, as shown  numerically in Ref. \cite{BrHaSi93},  the
dual London equation is satisfied in the MaA projection of
$SU(2)$ gluodynamics. A  recent detailed
investigation \cite{bali1,bali2} of the electric field profiles and
the distribution of monopole currents around the string shows that
the structure of the chromo-electric string in the MaA projection is
very similar to that of the Abrikosov string in the superconductor.

The following physical question is relevant: what  kind of
superconductor do we have. If $m_{ph} < m_{mon}$,  then the
superconductor is of the second type, the Abrikosov vortices  are
attracted to  each other. If $m_{ph} > m_{mon}$,  then the
superconductor is of the first type and the vortices repel each
other. The computation \cite{BrHaSi93} of the dual photon mass from
 the dual London equation shows that $m_{ph} \approx m_{mon}$. This
means that the vacuum of gluodynamics is close to the border between
type-I and type-II dual superconductors. The same conclusion is also
obtained in Ref. \cite{MaEjSu94}. A recent detailed study of the
abelian flux tube \cite{bali2} shows that the dual photon mass
is definitely smaller then the mass of the monopole, and therefore
the vacuum of $SU(2)$ is the type-II dual superconductor (see
also Ref. \cite{TS}).

\subsection{Fact~3: Monopole Condensate}

If the vacuum of $SU(2)$ gluodynamics in the abelian projection is
similar to the dual superconductor, then the value of the monopole
condensate should depend on the temperature as a disorder parameter:
at low temperatures it should be nonzero, and it should vanish above the
deconfinement phase transition.

The behaviour of the monopole condensate can be studied with the help
of the monopole creation operator. In gluodynamics this operator can
be derived by means of  the Fr\"ohlich and Marchetti construction
 \cite{FrMa86} for the compact electrodynamics. For $SU(2)$
lattice gluodynamics in the MA projection, it is convenient to study
the effective constraint potential for the monopole creation operator
(similar calculations were performed for compact electrodynamics
in Ref. \cite{PoPoWi91}):
\beqn
V_{eff}(\Phi) =
-\ln (<\delta (\Phi - \frac{1}{V}\sum_x \Phi_{mon}(x))>)
\label{Veff}
\eeqn
where $\varphi$ is the monopole field. If this potential has a
Higgs-like form
\beqn
V_{eff}(\Phi) \propto \lambda ({|\Phi|}^2 - \Phi^2_c)^2\,,
\qquad \lambda, \Phi^2_c >0\,,
\eeqn
then a monopole condensate exists. The dual
superconductor picture predicts this behaviour of the effective
constraint potential in the confinement phase. In the deconfinement
phase the following form of the potential is expected:
\beqn
V_{eff}(\Phi) \propto m^2 {|\Phi|}^2 +\lambda {|\Phi|}^4
+ \dots\,, \qquad m^2,\lambda > 0 \,;
\eeqn
(no monopole condensate).

\begin{figure}[htb]
\vskip5mm
\leftline{
\begin{tabular}{ll}
\hspace{-3mm}{\epsfxsize=0.48\textwidth\epsfbox{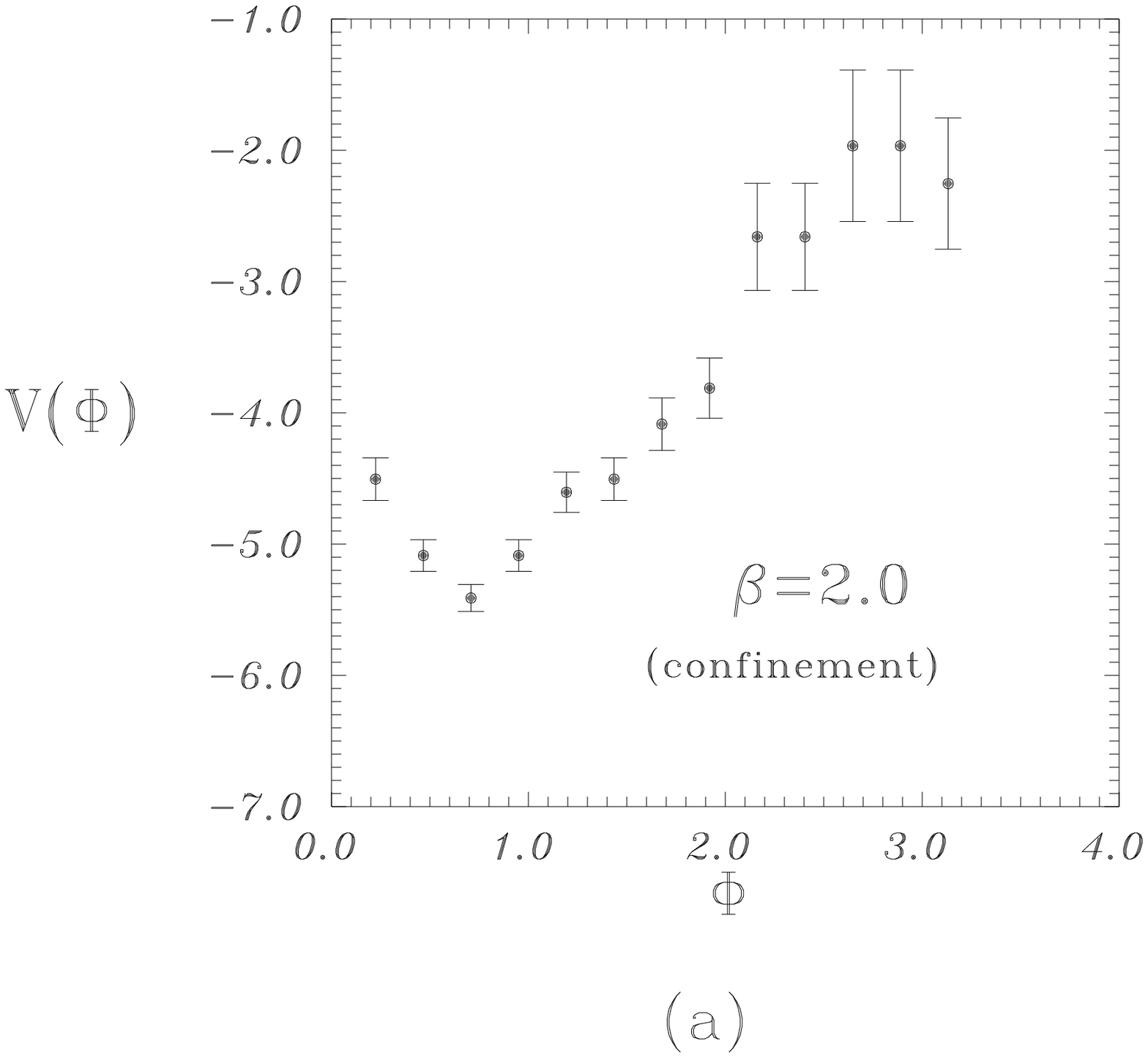}}&
{\epsfxsize=0.48\textwidth\epsfbox{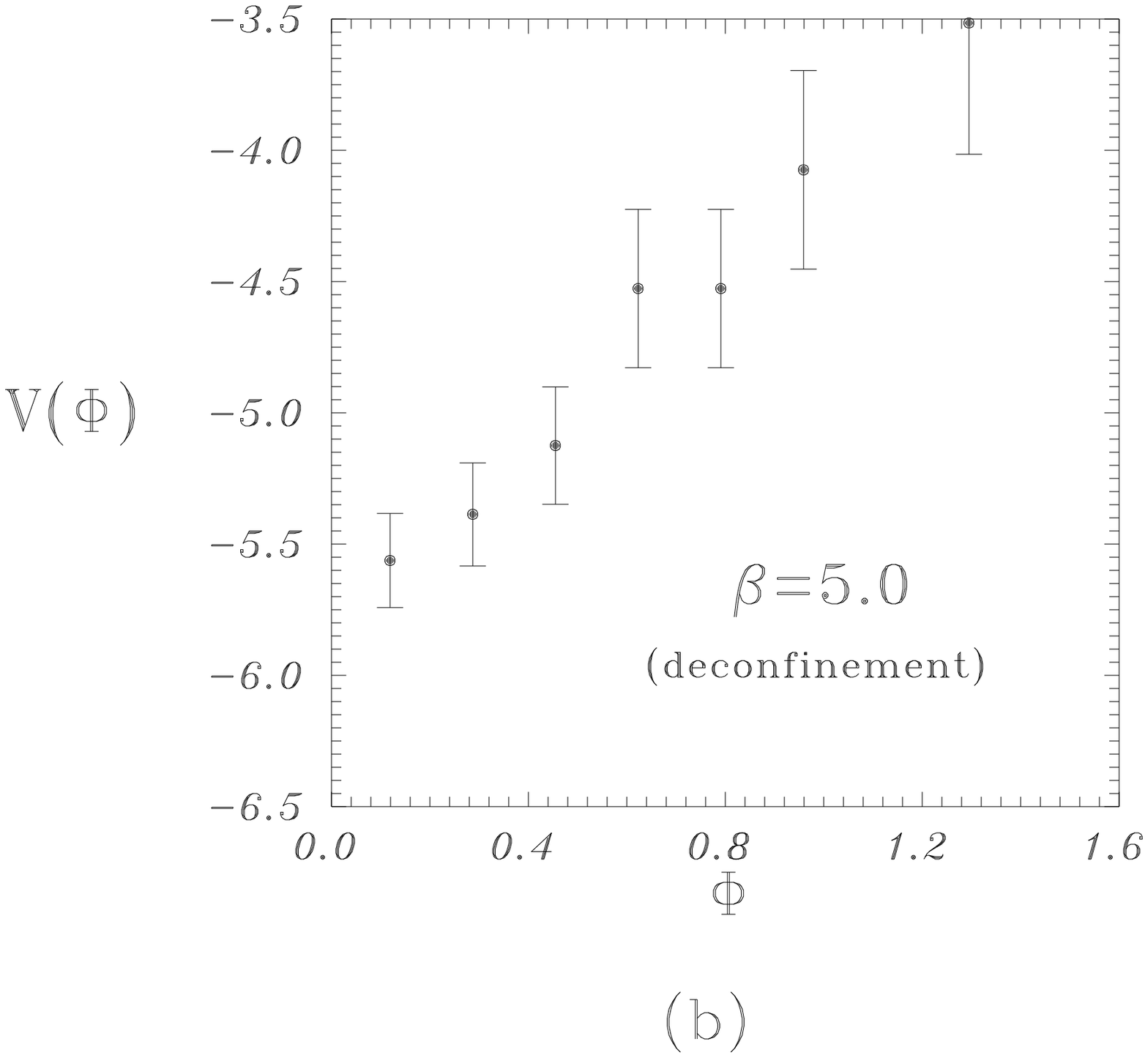}}\\
~ & ~\\
\end{tabular}
}
\vskip-7mm
\caption{
$V(\Phi)$ for the confinement~(a) and the deconfinement~(b) phases,
Ref. \cite{ChPoVe97}.}
\label{onefig}
\end{figure}
\begin{figure}[htb]
\vskip5mm
\leftline{
\begin{tabular}{cc}
{\epsfxsize=0.45\textwidth\epsfbox{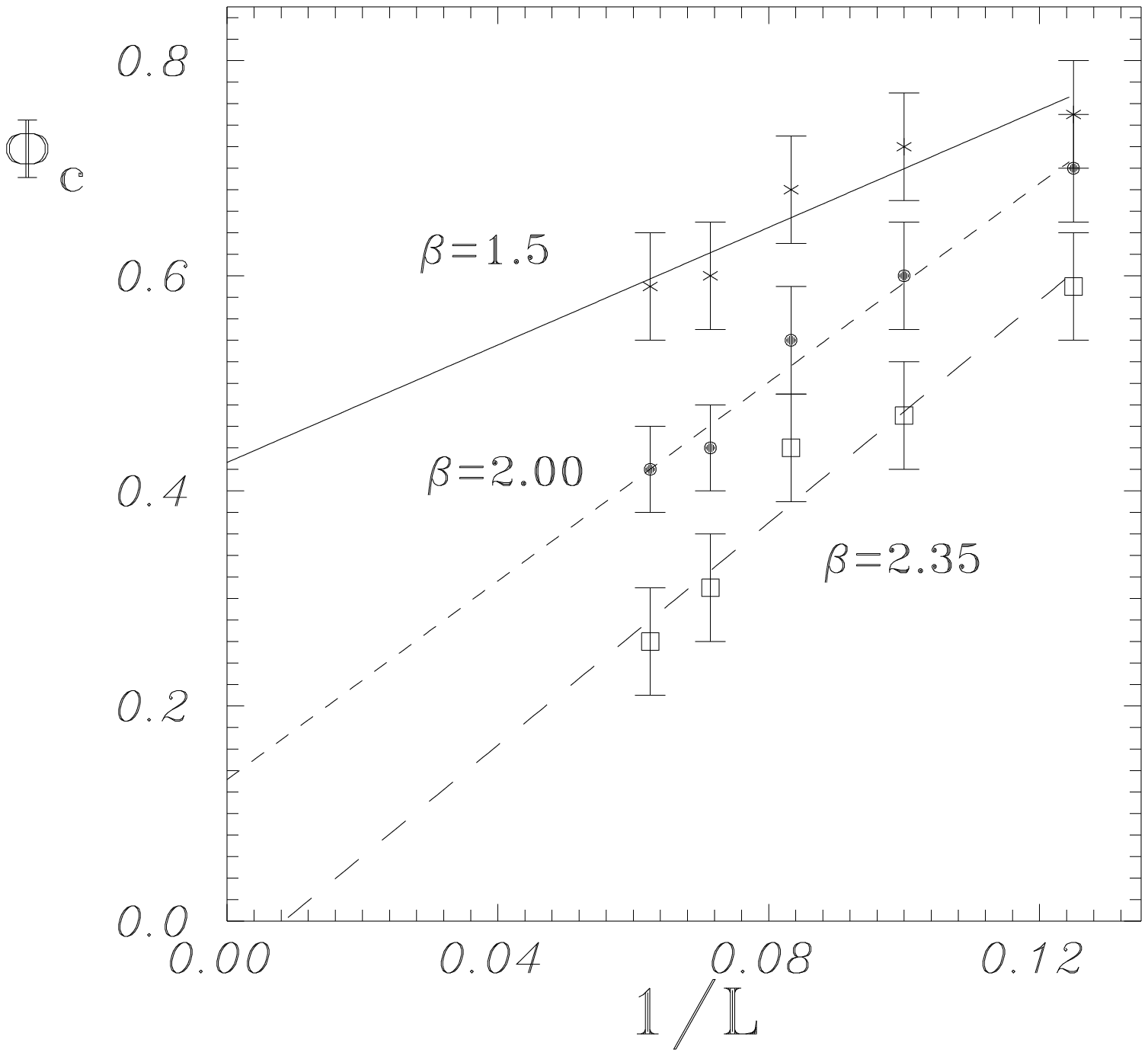}}&
{\epsfxsize=0.475\textwidth\epsfbox{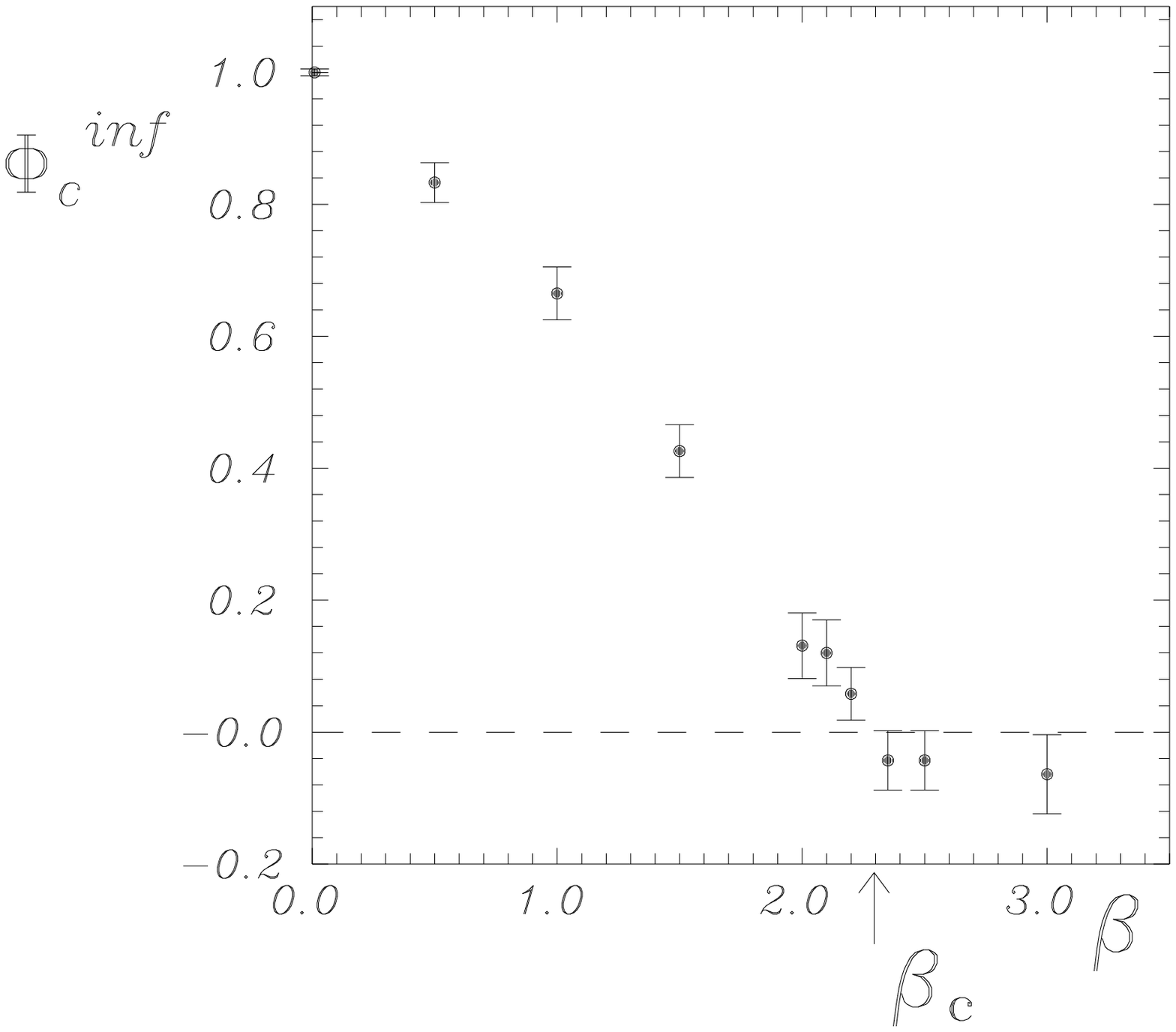}}\\
(a) & (b)\\
\end{tabular}
}
\caption{The dependence of (a) $\Phi_c$ on the spatial size of the lattice for
three values of $\beta$ and (b) of $\Phi_c^{inf}$ on $\beta$,
Ref. \cite{ChPoVe97}.}
\label{ldepend}
\end{figure}

The numerical calculation of the effective constraint potential
\eq{Veff} is very time--consuming, and  therefore, in
Refs. \cite{ChPoVe96,ChPoVe97} the following quantity $V(\Phi)$ has
been studied:
\beq
 V(\Phi) = - \ln (<\delta (\Phi - \Phi_{mon}(x))>)\,.
 \label{V}
\eeq
Numerical calculations  of this quantity were performed on lattices of
size $4\cdot L^3$, for $L = 8,10,12,14,16$ with anti--periodic
boundary conditions in space directions for the abelian fields.
Periodic boundary conditions are forbidden, since a $single$
magnetic charge cannot physically exist  in a  closed volume with
periodic boundary conditions due to the Gauss law.  The results on
 finite lattices have been extrapolated to the infinite volume,
since near the phase transition finite volume effects are very
strong.

In Figures~\ref{onefig} (a,b) the (right-hand side of the) effective
potential \eq{V} is shown for the confinement and the deconfinement
phases, the calculations being performed on a $4\cdot12^3$ lattice.

In the confinement phase (Figure~\ref{onefig} (a)), the minimum of
$V(\Phi)$ is shifted from zero, while in the deconfinement phase, the
minimum is at the zero value of the monopole field $\Phi$. The value
$\Phi_c$ of the monopole field  at which the potential has a minimum
is equal to the value of the monopole condensate.

The potential shown in Figure~\ref{onefig} (b) corresponds to a
trivial potential with a minimum at a zero value of the field:
$\Phi_c = 0$. The dependence of $\Phi_c$, the minimum of the potential,
 on the spatial size  $L$ of the lattice  is shown in
Figure~\ref{ldepend}(a).

We fit the data for $\Phi_c$ by the formula $\Phi_c = A L^\alpha +
\Phi_c^{inf}$, where $A$, $\alpha$ and $\Phi_c^{inf}$ are the fitting
parameters. It occurs that $\alpha = -1$ within statistical errors.
Figure~\ref{ldepend}(b) shows the dependence on $\beta$ of the value of the
monopole condensate  extrapolated to the infinite spatial volume
$\Phi_c^{inf}$. It is clearly seen that $\Phi_c^{inf}$ vanishes at
the point of the phase transition and it plays the role of the order
parameter.

The monopole creation operator \cite{KeKi86} in the monopole current
representation is studied in Ref. \cite{Nak96}.  First the monopole
action is reconstructed from the monopole currents in the MA
projection, and after that the expectation value of the monopole
creation operator is calculated in the quantum theory of monopole
currents. Again, the monopole creation operator depends on the
temperature as the disorder parameter. A slightly different monopole
creation operator was studied in Refs. \cite{DG,DiGi}.

\subsection{Abelian Monopole Action: Analytical examples}

In the next section we explain how to study the action of the
monopoles extracted from $SU(2)$ fields in the MaA projection. Now we
give several examples of monopole actions corresponding to abelian
gauge theories.

\fitem{Example~1:}{$4D$ compact abelian electrodynamics}

We have already discussed  the duality transformation of $4D$ compact
electrodynamics. There is  another exact transformation which
represents the partition function as a sum over monopole
currents. This transformation was initially  performed by Berezinsky
 \cite{Ber} and by Kosterlitz and Thouless \cite{KoTh} for the $2D$ $XY$
model.  Accordingly, we  call it  the $BKT$ transformation.
For compact electrodynamics with the Villain action this
transformation was found by Banks, Myerson and Kogut \cite{BMK}. The
partition function for compact electrodynamics with the Villain
action is
\beqn
 \cZ^{Villain} = \intpi \dD \theta
 \nsum{m}{2} \exp\Bigl\{- {||{\rm d} \theta + 2 \pi m||}^2 \Bigr\}\,.
 \label{Villain}
\eeqn
As explained in  Appendix~C,  the BKT transformation of this partition
function has the form
\beqn
\cZ = [\int\limits_{-\infty}^{+\infty} \cD A \exp
\{-\beta {||\dd A||}^2 \}] \times
[\nddsum{j}{1} \exp\{ - 4\pi^2 \beta \,
(\dual j, \Delta^{-1} \dual j)\}] \,.
\eeqn
Here the partition function for the non--compact gauge field $A$
is  inessential because  it is  Gaussian. All the  dynamics  is in the
partition function for the monopole currents $\dual j$ which are lying
on the dual lattice and form  closed loops ($\delta \dual j = 0$).

It is possible to find the BKT transformation for compact
electrodynamics with a general form for the action \cite{ChPoVe97}. In
this general case the monopole action is non-local (see Appendix~C).

\fitem{Example~2:}{Abelian Higgs theory in the London limit}

The partition function for the Abelian Higgs model is (cf.,
eqs.(\ref{PAH},\ref{SAH}\footnote{We assume the Villain form for the
interaction of the gauge and Higgs fields.}):
\beqn
\cZ^{AH} =
\int\limits^{+\infty}_{-\infty} \cD B \int\limits^{+\pi}_{-\pi}
\cD \varphi\, \nsum{m}{2}
\exp\Bigl\{ - \beta {||\dd B ||}^2 - \gamma
{||\dd \varphi + B + 2 \pi m||}^2 \Bigr\} \,,
\eeqn
where $B$ is the non-compact gauge field and $\varphi$ is the phase
of the Higgs field (the radial part of the Higgs field is frozen,
since we consider the London limit).

After the BKT transformation
 \cite{IvPo91,SmSi91}, the partition function takes the form
\beqn
\cZ^{AH} = \const\!\!\!\!\! \nDsum{j}{1} \exp \Bigl\{
- S^{AH}_{mon}(j) \Bigr\}\,, \quad
S^{AH}_{mon}(j) = \frac{1}{4 \beta} (j, \Delta^{-1} j)
+ \frac{1}{4 \gamma} {|| j ||}^2\,,
\label{star}
\eeqn
where $\Delta^{-1}$ is the inverse lattice Laplacian and $j$ is the
current of the Higgs particles.

\fitem{Example~3:}{Abelian Higgs theory near the London limit}

As pointed out by T.~Suzuki \cite{TS},  the numerical
data show that in the MaA projection the currents in lattice
$SU(2)$ gluodynamics behave as the currents of the Higgs particles in
the Abelian Higgs model near the London limit. This  means that the
coefficient $\lambda$ of the Higgs potential $\lambda
{({|\Phi|}^2-1)}^2$, eq.\eq{SAH}, is large but finite. In this
case it is impossible to get an explicit expression for the monopole
action, but there exists a $1 \slash \lambda$ expansion of this
action
\beqn
S_{mon} (\dual j) & = &
\frac{1}{4 \beta} (\dual j,\Delta^{-1} \dual j) +
\Bigl(\frac{1}{4 \gamma} + a_1 \Biggr) {||\dual j ||}^2 + a_2
\sum\limits_x {\Bigl(\sum\limits^4_{\mu=-4} \dual j^2_{x,\mu}
\Biggr)}^2 \nonumber\\
& & + a_3 \sum\limits_x \sum\limits^4_{\mu=1}
\dual j^4_{x,\mu} + \dots\,,
\label{starstar}
\eeqn
where
\beqn
a_k = \sum^{\infty}_{n=1} \frac{f^{(n)}_k}{\lambda^n}\,, \nonumber
\eeqn
$f^{(n)}_k$ are the coefficients which can be
calculated \cite{TS}. In the limit $\lambda \to + \infty$,
this action is reduced to $S^{AH}_{mon}(\dual j)$, eq.\eq{star}.

\subsection{Monopole Action from ``Inverse Monte--Carlo"}

The $1 \slash \lambda$ expansion of the monopole action has an
important application in lattice gluodynamics. It is possible to show
that the monopole currents in the MaA projection of $SU(2)$
gluodynamics are in a sense equivalent to the currents generated by
the theory with the action \eq{starstar}. This means that (at least at
large distances) gluodynamics is equivalent to the Abelian Higgs
model. The details can be found  in the lecture of T.~Suzuki
 \cite{TS}. Below we briefly describe some of these results.

It occurs that from a given distribution of currents on the $4D$
lattice it is possible to find the action of the currents. This can
be done by the Swendsen ("inverse Monte--Carlo")
 \cite{Sw,Suzuki9-10} method. By the usual Monte--Carlo method we
get an ensemble of the fields $\{\Phi\}$ with the probability
distribution $e^{-S(\Phi)}$. The "inverse Monte--Carlo" allows us  to
reconstruct the action $S(\Phi)$ from the distribution of the fields
$\{\Phi\}$.

\begin{figure}[htb]
\vskip5mm
\begin{center}
\begin{tabular}{cc}
{\epsfxsize=0.45\textwidth\epsfbox{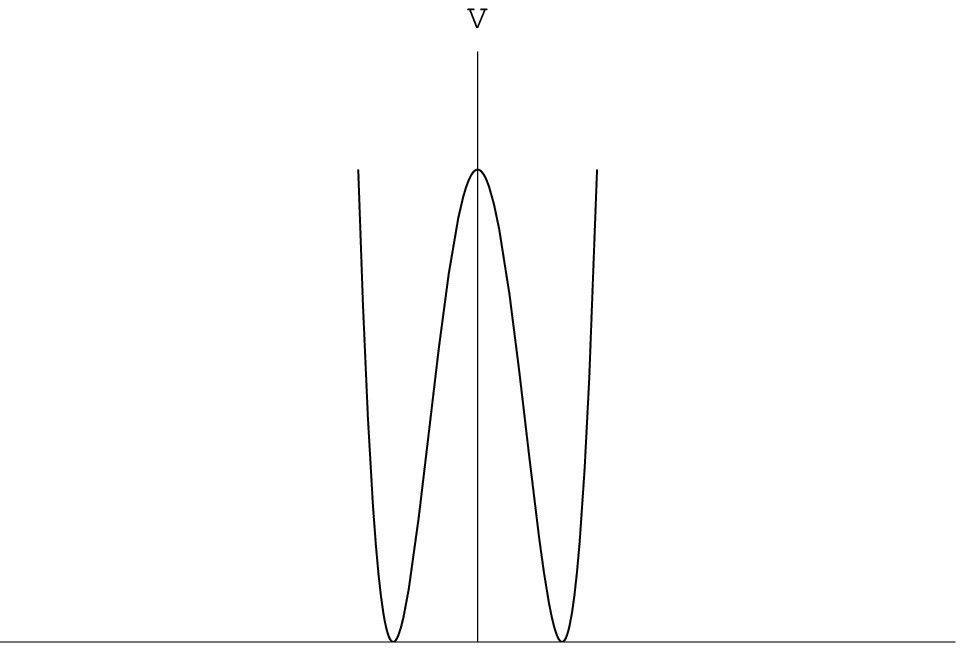}}&
{\epsfxsize=0.45\textwidth\epsfbox{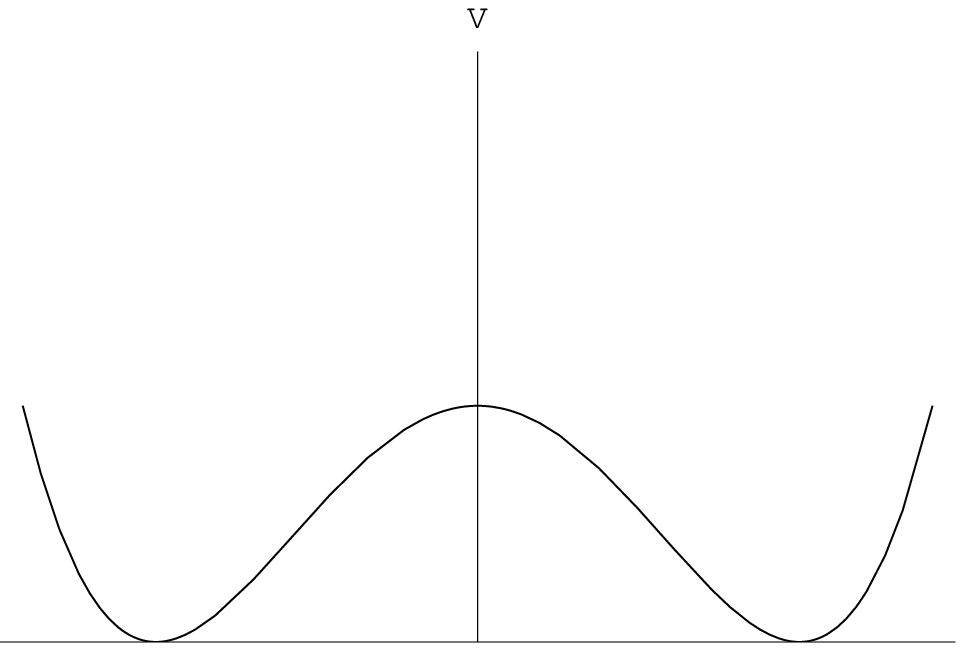}}\\
(a) & (b)\\
\end{tabular}
\end{center}
\vskip-5mm
\caption{Schematic representation of the monopole potential $\lambda
{({|\Phi|}^2 - 1)}^2$ for compact electrodynamics (a), $\lambda
\to + \infty$; and for lattice $SU(2)$ gluodynamics (b), $\lambda$ is
finite.}
\label{vv}
\end{figure}

The procedure of the reconstruction of the monopole action for
lattice gluodynamics is the following. First the $SU(2)$ gauge fields
are generated  by the  usual Monte--Carlo  method. Then, the MaA gauge
fixing is performed and the abelian monopole currents are extracted
from the abelian gauge fields. Finally,  the inverse Monte--Carlo
method is applied to the ensemble of these currents and the
coefficients  $\beta,\gamma,a_1,a_2,a_3,\dots$ of the action \eq{starstar}
are determined. It occurs that at large distances the distribution of
the currents is well described by the action \eq{starstar}. In  that
sense,  the effective action of gluodynamics is the Abelian Higgs
model, the monopoles play the role of the Higgs particles. In
contrast to compact electrodynamics,  the monopole potential is not
infinitely deep, see Figure~\ref{vv}(a,b).

\subsection{Various Representations for the Partition Function}

There are several equivalent representations  of the partition
function:
\beqn
\cZ^{j} = \sum\limits_{\delta j=0} \exp\{- S_{mon} (j)\}\,,
\label{MonPF}
\eeqn
where $S_{mon}$ is given by \eq{starstar}.

\fitem{Representation~1}{as the dual abelian Higgs model:}

As we have already said, the  model
\eq{MonPF} is related to the (dual) Abelian Higgs theory by the
inverse BKT transformation
\beqn
\cZ^{j} \propto \cZ^{AH} & = & \int \cD B \, \cD \Phi
\exp \{- S^{AH}(B,\Phi)\}\,,
\label{AH}
\eeqn
where $S^{AH}$ is given by eq.\eq{SAH}.

\fitem{Representation~2}{in terms of the Nielsen-Olesen strings:}

Using the BKT transformation for the partition function \eq{AH}, one
can show that model \eq{MonPF} is equivalent \cite{PoWiZu93} to the
string theory:
\beqn
\cZ^{j} \propto \cZ^{str} & = & \nDsum{\sigma}{2}
e^{-\cS^{AH}_{string}(\sigma)} \label{11} \\
\cS^{AH}_{string}(\sigma) & = & 4 \pi \gamma (\sigma,
{(\Delta + m^2)}^{-1} \sigma) { + O(\frac 1\lambda)}\,,\quad
m^2 = \frac{\gamma}{\beta}\,.\label{12}
\label{StrMon}
\eeqn
Here $\sigma$ is the closed world sheet of the Nielsen-Olesen string.
In the continuum limit \cite{Orland} we have
\beqn
\cS^{AH}_{string} \propto
{\int} d \sigma_{\mu \nu}(\tilde x)
\cD^{(4)}_m(\tilde x - \tilde x')d \sigma_{\mu \nu}(\tilde x')\,,
\eeqn
where $\cD^{(4)}_m$ is the free scalar propagator, $(\Delta+m^2)
\cD^{(4)}_m(x)=\delta^{(4)}(x)$.

\fitem{Representation~3}{in terms of the gauge field $\theta$:}

Using the inverse dual transformation for the partition
function \eq{AH}, one  can get the representation of the
monopole partition function in terms of the $compact$ gauge field
$\theta$ which is dual to the gauge field $B$, \eq{SAH}, \eq{AH}:
\beqn
\cZ^{j} \propto \cZ^{\theta} & = & \int\limits_{-\pi}^{+\pi}
\nsum{n}{2} \cD \theta e^{-\cS^{o}(\theta,n)}\,;
\label{9}\\
\cS^{o} & = & \frac{1}{16 \pi^2 \beta}
{||\dd \theta + 2 \pi n ||}^2 -
\frac{1}{16 \pi^2 \gamma} \left(
\dd \theta + 2 \pi n, \Delta (\dd \theta + 2 \pi n)  \right)
+ O(\frac 1\lambda)\,.
\eeqn

\fitem{Representation~4}{in terms of hypergauge fields:}

Applying the duality transformation to the partition function \eq{9},
we get the representation in terms of the hypergauge (Kalb--Ramond)
field
\beqn
\cZ^j & \propto & \cZ^{HG} = \nsum{h}{2}
{\int} \cD A \exp\{-\cS(A,{ h})\} \nonumber\\
\cS(A,h) & = & \frac{1}{4 \gamma} \|d { h} \|^2 +
\frac{1}{4 \beta} \| d A + { h}\|^2 + O(\frac 1\lambda) \,,
\eeqn
where $h$ is the hypergauge field ($h_{\mu\nu}(x)$) interacting
with the gauge field
$A$ ($A_\mu(x)$); the field $h$ is dual to the monopole field $\Phi$,
and the field $A$ is dual to the field $B$.

The action $\cS(A,h)$ is invariant under gauge transformations:
$A \to A + \dd \alpha$, $h \to h$,  and under {\it hypergauge}
transformations $A \to A - \gamma$, $h \to h + \dd \gamma$.

\fitem{Representation~5}{by fourier transformation of the string world
sheets:}

This representation is intermediate between the representations
\eq{11} and \eq{AH}:
\beqn
\cZ \propto \cZ^h & = & \int \cD \xi \int \cD \zeta
\nsum{\sigma}{2} \, \exp\left\{ 2\pi i  (\xi, \sigma)\, - \,
\tilde S [\xi,\zeta] + O(\frac 1\lambda) \right\}\,,\\
S[\xi,\zeta] & = & \frac{1}{4 \beta} {|| \xi + \dd \zeta||}^2+
\frac{1}{4 \gamma} {|| \dd \xi||}^2\,,
\eeqn
where $\sigma$ are the world sheets of the Abrikosov-Nielsen-Olesen
strings. Again,  there exists a hypergauge symmetry ($\xi \to \xi +
\dd \gamma$, $\zeta \to \zeta - \gamma$), which reflects the
fact that the string worlds sheets are closed (or equivalently, conservation
of the magnetic flux).

It is possible to derive a lot of physical consequences from these
representations \cite{TS}. One of the most interesting is
that the classical string tension which is calculated from the string
action \eq{12} coincides, within statistical errors,  with the
string tension in $SU(2)$ lattice gauge theory. The coefficients
$\beta$ and $\gamma$ in eq.\eq{12} are   found by the inverse
Monte--Carlo method,  as we have  just explained.

\subsection{Abelian Monopoles as Physical Objects}

The abelian monopoles arise in the continuum theory \cite{tH81} from
singular gauge transformation \eq{8a}-\eq{mcharge} and it is not
clear whether these monopoles are ``real'' objects. A physical object
is something which carries action and below we only  discuss the
question if there are any correlations between abelian monopole
currents and the $SU(2)$ action. In Ref. \cite{Suzuki9-10} it was
found that the total action of the $SU(2)$ fields is correlated with
the total length of the monopole currents, so there exists a global
correlation. We now discuss the local correlations between the action
density and the monopole currents \cite{ActMonCorr}.

In lattice calculations,  the monopole current $j_\mu (x)$ lies on the
dual lattice and it is natural to consider the correlator of the
current and the dual action density\footnote{Here and below,
formulae correspond to the continuum limit implied by the
lattice regularization.}:
\beqn
C_1 = <\frac 12 Tr\,\left( j_\mu (x)\,
\frac{1}{2} \varepsilon_{\nu\mu\alpha\beta}
F_{\alpha\beta}(x)\right)^2 > - <j_\mu^2 (x)> <\frac 12 Tr
F_{\alpha\beta}^2(x)> \, .
\eeqn
The density of $j_\mu (x)$ strongly depends on $\beta =
\frac{4}{g^2}$, therefore it is convenient to normalize $C_1$:
\beqn
C_2 = \frac{<\frac 12 {\rm Tr} \,\left( j_\mu (x)\,
\frac{1}{2} \varepsilon_{\nu\mu\alpha\beta}
F_{\alpha\beta}(x)\right)^2 >}{<j_\mu^2 (x)>
<\frac 12 {\rm Tr} F_{\alpha\beta}^2(x)>} - 1 \, .
\eeqn
For the static monopole we have $j_0 (x) \neq 0$, $j_i (x) =0, \,
i=1,2,3$,  and the fact that $C_2 \neq 0$ means that the magnetic part
$S_m$ of the $SU(2)$ action is correlated with $j_\mu (x)$.

The correlator $C_2$ is related to the
relative excess of the action carried by the monopole current. The
expectation value of $S_m$ on the monopole
current is
\beqn
S_m = <\frac {1}{24} {\rm Tr} \,\left( n_\mu (x)\,
\frac{1}{2} \varepsilon_{\nu\mu\alpha\beta}
F_{\alpha\beta}(x)\right)^2 > \, ,
\eeqn
where $n_\mu (x)=j_\mu (x) \slash | j_\mu (x) |$. The
relative excess of the action on the monopole current is
\beqn
\eta = \frac{S_m - S}{S}\,,\quad
S = <\frac {1}{24} Tr F_{\alpha\beta}^2(x)>\,.
\label{etaF}
\eeqn
where $S = <\frac {1}{24} Tr F_{\alpha\beta}^2(x)>$ is the expectation
value of the action, the coefficient $\frac{1}{24}$ corresponds to
the lattice definition of the ``plaquette action'' $<S>$. If $j_\mu
(x) = 0, \pm 1$,  then $\eta = C_2$. Since in lattice calculations at
sufficiently large values of $\beta$ the probability of $j_\mu (x) =
\pm 2$ is small in the MaA projection, we have  $\eta \approx C_2$ at large
values of $\beta$. From numerical calculations we have found that $\eta =
C_2$ with  5\%  accuracy for $\beta > 1.5$ on lattices of sizes
$10^4$ and $12^3 \cdot 4$. The lattice definition of $S_m$ is
\beq
S_m= <\sum^4_{\nu=1} n_\nu^2 (x) \cdot \frac 16
\sum_{P \in \partial C_\nu (x)}
\left( 1 - \frac 12 Tr \, U_P \right)>\, , \label{Smlat}
\eeq
where the summation is over the plaquettes $P$ which are the faces of
the cube $C_\nu (x)$; the cube $C_\nu (x)$ is dual to $n_\nu(x)$;
$U_P$ is the plaquette matrix.
Thus,  $S_m$ is the average action on the plaquettes  closest
to the magnetic current (see Figure~\ref{monopolecurrent}(b)). The
lattice definition of $S$ is standard: $S = <\left( 1 - \frac 12 Tr
\, U_{P}\right)>$. We use these definitions of $S$ and $S_m$ in our
lattice calculations.

\begin{figure}[t!]
\vskip5mm
\begin{center}
\begin{tabular}{cc}
\hspace{-3mm}{\epsfxsize=0.48\textwidth\epsfbox{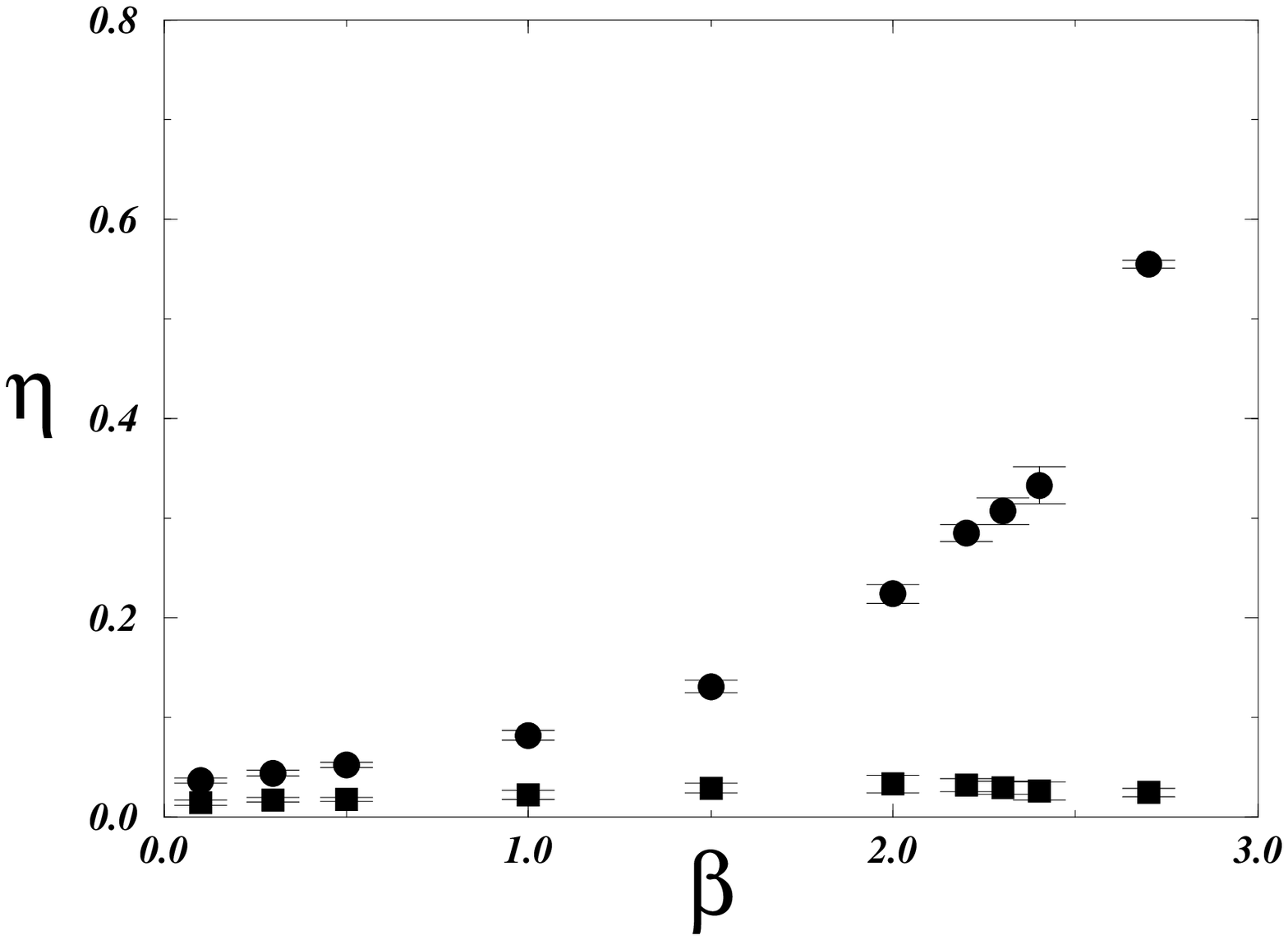}}&
{\epsfxsize=0.48\textwidth\epsfbox{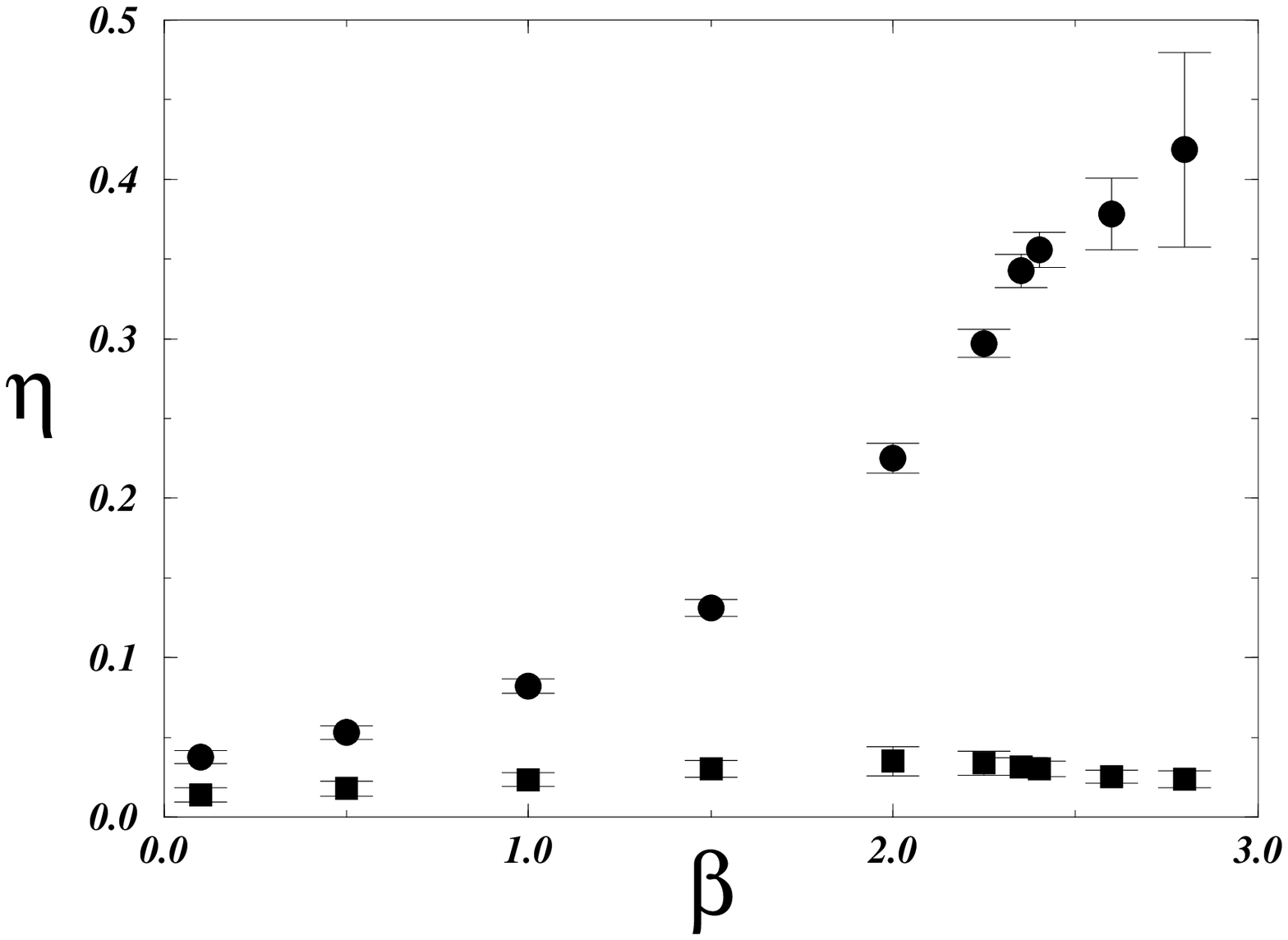}}\\
 (a) & (b)\\
\end{tabular}
\vskip-5mm
\end{center}
\caption{The relative excess $\eta$, eq.\eq{etaF}, for $10^4$ lattice (a)
and for $12^3 \cdot 4$ lattice (b) in
the MA (circles) and the Polyakov abelian (squares) gauges,
Ref. \cite{ActMonCorr}.}
\label{etas}
\end{figure}

It is interesting to study the quantity \eq{etaF} in various gauges.
In Figure~\ref{etas}(a) the relative excess of the magnetic action
density near the monopole current  $\eta$  is shown for a $10^4$
lattice. Circles correspond to the MaA projection,  and squares
to the  Polyakov gauge. The data for the $F_{12}$ projection coincide,
within statistical errors,  with those  for the Polyakov gauge.

The quantity \eq{etaF} at  finite temperatures  has  also been studied.
In Figure~\ref{etas}(b) the relative excess of the magnetic action
density near the monopole current  $\eta$  is shown for $12^3\cdot 4$
lattice. Again, the circles correspond to the MaA projection and the
squares  to the  Polyakov gauge.

Thus we have shown that in the MaA projection the {\it abelian}
monopole currents are surrounded by regions with a high {\it
nonabelian} action. This fact presumably means that the monopoles in the MaA
projection are physical objects. It does not mean that they have to
be real objects in the Minkowsky space. What we have  found is that
these currents carry $SU(2)$ action in Euclidean space. It is
important to understand what is the general class of configurations
of $SU(2)$ fields which generate the monopole currents. Some specific
examples are known. These are instantons \cite{InstMonKucha} and the
BPS--monopoles (periodic instantons) \cite{SmSi91}.

\subsection{Monopoles are Dyons}

A dyon is an object which has both electric and magnetic charge.
In the field of a  single instanton
the monopole currents in the MaA  projection are accompanied
by electric currents \cite{BoSc}. The qualitative
explanation of this fact is simple. Consider the (anti)self-dual
configuration
\beq                 \label{duality}
F_{\mu\nu}(A)=\pm\dual
F_{\mu\nu}(A)\;.
\eeq
The MaA projection is defined \cite{KrScWi87} by the minimization of
the off-diagonal components of the non-abelian gauge field, so that
in the MaA gauge one can expect the  abelian component of the
commutator term
$1/2Tr(\sigma^3[A_{\mu},A_{\nu}])=\varepsilon^{3ab}A^a_{\mu}A^b_{\nu}$ to be  small compared with the abelian field-strength
$f_{\mu\nu}(A)=\diff_{[\mu}A^3_{\nu ]}$.
Therefore, in the MaA projection eq.(\ref{duality}) yields
\beq                                          \label{abelian_duality}
f_{\mu\nu}(A)\approx\pm\dual f_{\mu\nu}(A)\,.
\eeq
Due to eq.(\ref{abelian_duality}),  the monopole currents have to be
correlated with electric ones,  since
\beq
\label{jm_je}
j^e_{\mu}=\diff_{\nu} f_{\mu\nu}(A)\approx
\pm\diff_{\nu}\dual f_{\mu\nu}(A)=j^m_{\mu}\;.
\eeq

\begin{figure}[b!]
\vskip5mm
\begin{center}
\begin{tabular}{cc}
{\epsfxsize=0.4\textwidth\epsfbox{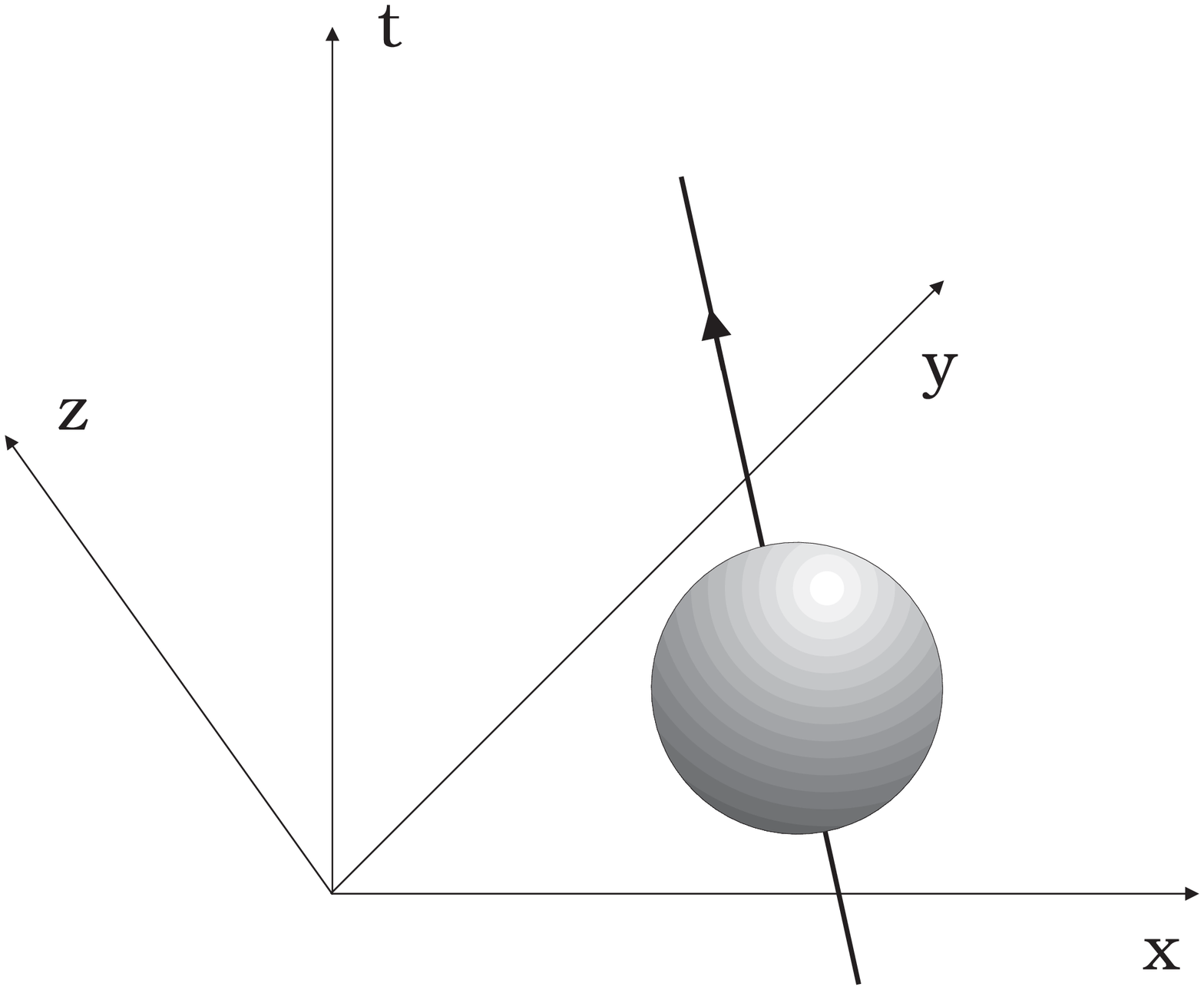}}&
{\epsfxsize=0.4\textwidth\epsfbox{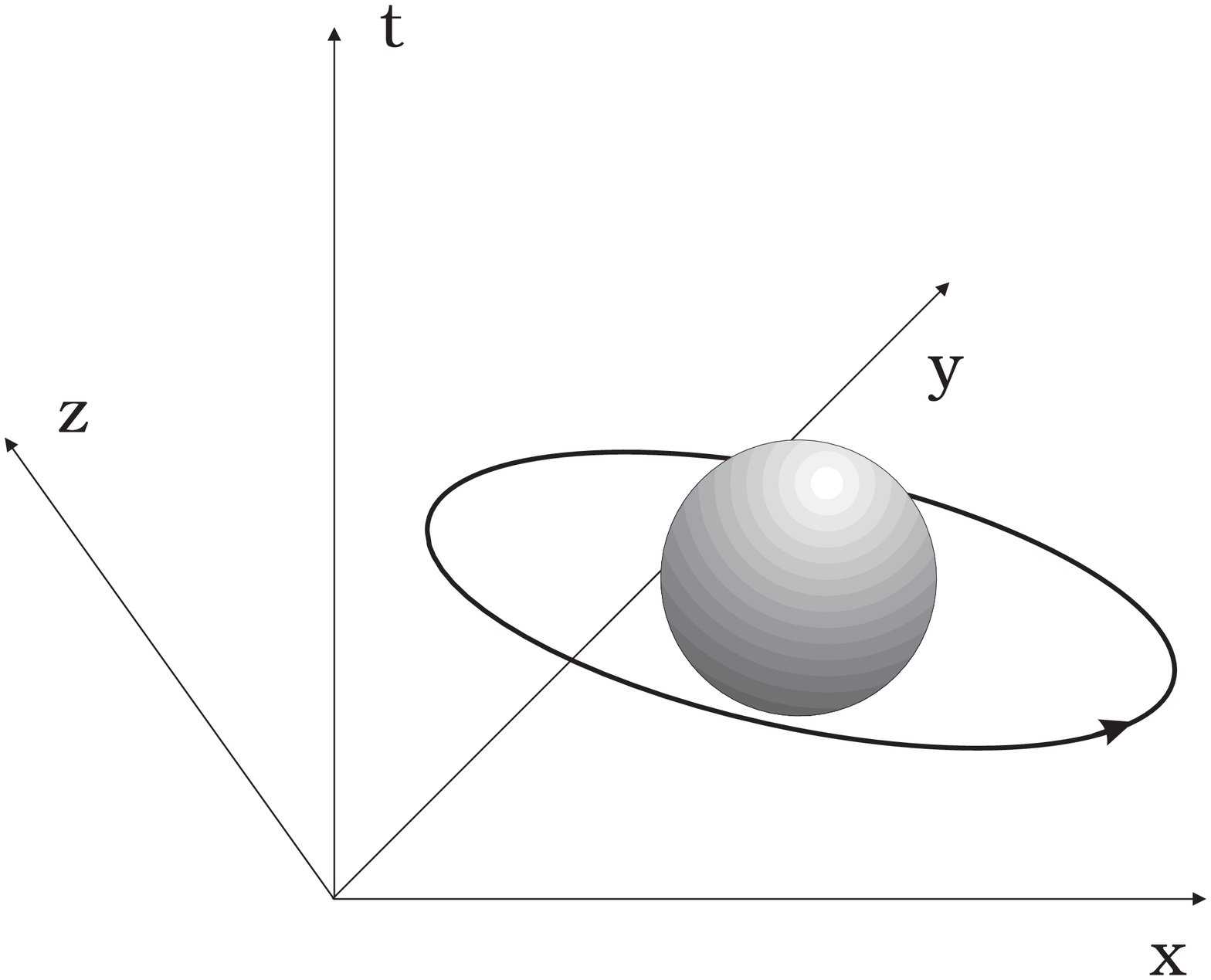}}\\
(a) & (b)\\
\end{tabular}
\end{center}
\vskip-5mm
\caption{The instantons (spheres) inducing monopoles (thin
lines).}
\label{inst}
\end{figure}

It is known that the (anti-) instantons produce abelian monopole
currents \cite{InstMonKucha}. The abelian monopole trajectories may
go through the center of the instanton (Figure~\ref{inst}(a)) or form
a circle around it (Figure~\ref{inst}(b)).

Due to eq.\eq{jm_je},  the monopole currents are accompanied by
electric currents in the self-dual fields. Therefore,  the monopoles
are dyons for an instanton background. However, the real vacuum is
not an ensemble of instantons,  and below we discuss the
correlation of $j^e$ and $j^m$ in the vacuum of lattice gluodynamics.

There is a  lot  of  vacuum models: the instanton gas, instanton
liquid, toron liquid,  {\it etc}.  Suppose that the vacuum is a
``topological medium'', and  there are self-dual and anti-self-dual regions
(domains). In this vacuum the electric and the magnetic currents should
correlate with each other. But the sign of the correlator depends on
the topological charge of the domain.

We define the electric current as \cite{BoSc}
\beq
\label{electric_current}
J^e_{\mu}(x)=\frac{1}{2\pi} \sum\limits_{\nu}
[\bar\theta_{\mu\nu}(x)-\bar\theta_{\mu\nu}(x-\hat{\nu})].
\eeq
In the continuum limit, this definition
corresponds to the usual one: $J^e_{\mu}=\diff_{\nu} f_{\mu\nu}$.
The electric currents are conserved
($\diff_{\mu} J^e_{\mu}=0$) and are attached to the links  of the
original lattice.  Electric currents are not
quantized.

In order to calculate the correlators of the type
$\corr{J^e_{\mu}(x)J^m_{\mu}(x)...}$,
one has to define the electric current on the dual lattice
or the  magnetic current on the original lattice.
We define the electric current on the dual lattice in
the following way:
\beq                                         \label{correspondence}
J^e_{\mu}(y) = \frac{1}{16}\summ{x\in \dual C(y,\mu)}{} \left[
J^e_{\mu}(x)+J^e_{\mu}(x-\hat{\mu})\right]\;.
\eeq
Here  the  summation on the
r.h.s. is over eight vertices $x$ of the 3-dimensional cube $\dual C(y,\mu)$
to which the current $J^e_{\mu}(y)$ is dual. The point $y$ lies on the dual
lattice and the points $x$  on the original one.

For the topological charge density operator
we use the simplest definition:
\beqn
Q(x) = \frac{1}{2^9 \pi^2} \sum_{\mu_1,...,\mu_4=-4}^{4}
\varepsilon^{\mu_1,...,\mu_4}
Tr[U_{\mu_1,\mu_2}(x) U_{\mu_3,\mu_4}(x)],
\eeqn
where $U_{\mu_1,\mu_2}$ is the plaquette matrix. On the dual lattice
the topological charge density corresponding to the monopole current
$J^m_{\mu}(y)$ is defined by taking the average over the eight sites
nearest to the current $J^m_{\mu}(y)$: $Q(y)=\frac{1}{8}\summ{x}{}
Q(x)$.

The simplest (connected) correlator of electric and magnetic currents is
given by\\
 ${<J^m_\mu J^e_\mu >}_c \equiv <J^m_\mu J^e_\mu>
- <J^m_\mu> <J^e_\mu> $ $= <J^m_\mu J^e_\mu>$, here we used the fact
that $<J^m_\mu> = <J^e_\mu> =0$ due to the Lorentz invariance. The
connected correlator ${<J^m_\mu J^e_\mu >}_c$ is equal to zero, since
$J^m_{\mu}$ and $J^e_{\mu}$ have opposite parities. A scalar
quantity can be constructed if we multiply $J^m_{\mu}J^e_{\mu}$ with
the topological charge density. The corresponding connected
correlator
\beq
\label{JmJeQ}
{<J^m_\mu J^e_\mu Q>}_c = <J^m_\mu J^e_\mu Q>
\eeq
is nonzero for the vacuum consisting of (anti-)self-dual
domains~({\it cf.} eq.~(\ref{jm_je})). Note that we derived
eq.\eq{JmJeQ} using the equalities $<J^m_\mu> = <J^e_\mu> = <Q> = 0$.

We consider $SU(2)$ lattice gauge theory on a  $8^4$ lattice with
the Wilson action. We calculate \cite{jmje} the correlator
$\corr{J^m_{\mu}(y) J^e_{\mu}(y) Q(y)}$, using 100 statistically
independent configurations at each value of $\beta$.

This correlator strongly depends on $\beta$ and it is convenient to
normalize it  by dividing by $\rho^m \rho^e$. Here $\rho^m$ and $\rho^e$
are the monopole and the electric current  densities:
\beq
\rho_{m(e)}=\frac{1}{4V}\summ{l}{} |J^{m(e)}_l|,
\eeq
$V$ is the lattice volume (the total number of sites).

\begin{center}
\begin{figure}[htb]
\centerline{\epsfig{file=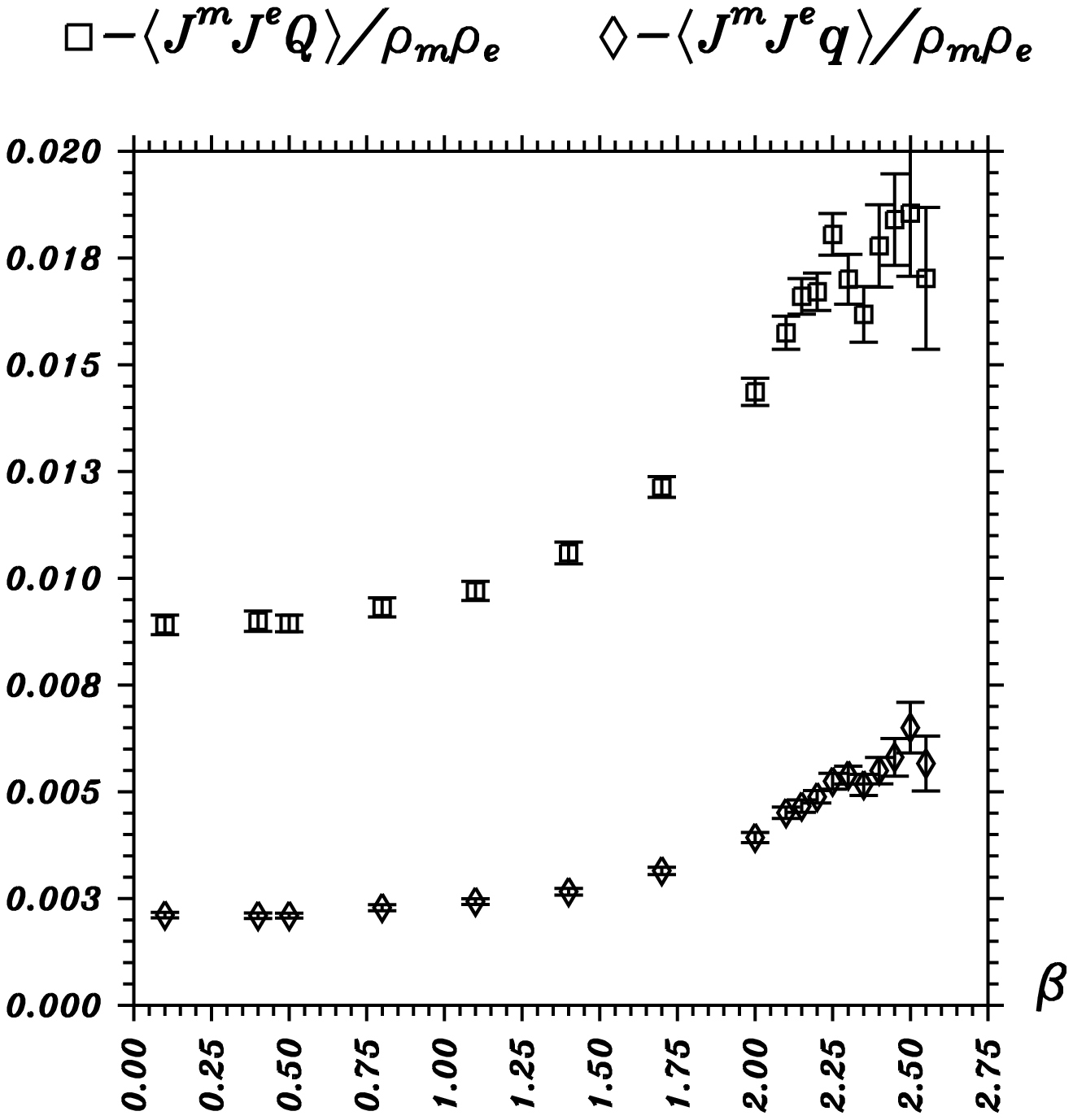,height=8cm}}
\caption{
Correlators $<\!\!J^m J^eQ\!\!>\!\!/\rho_m\rho_e$
and $<\!\!J^mJ^eq\!\!>\!\!/\rho_m\rho_e$
as  functions  of $\beta$, Ref. \cite{jmje}.}
\label{12jmje}
\end{figure}
\end{center}

The correlators $\corr{J^mJ^eQ}\!\!/\rho_m\rho_e$ and
$\corr{J^mJ^eq}\!\!/\rho_m\rho_e \quad (q(y)=Q(y)/|Q(y)|)$ are
represented in Figure~\ref{12jmje}. As one can see from
Figure~\ref{12jmje}, the product of the electric and the magnetic
currents is correlated with the topological charge.

Thus our results show that in the vacuum of lattice  gluodynamics the
magnetic current is correlated with the electric current, the
abelian monopoles have an electric charge. The sign of the electric
charge depends on the sign of the topological charge density.

\section{Conclusions}

Now we briefly summarize the properties of the abelian monopole
currents in the MaA projection of lattice $SU(2)$ gluodynamics:

\begin{itemize}
\item Monopoles are responsible for $\approx 90\%$
of the $SU(2)$ string tension.
\item Monopole currents satisfy the London equation
for a superconductor.
\item Monopoles are condensed in the confinement phase.
\item The effective monopole Lagrangian is similar to the Lagrangian
of the dual Abelian--Higgs model.
\item Monopoles carry the $SU(2)$ action.
\item Monopoles are dyons.
\end{itemize}

The main conclusion which can be obtained from these facts is that
the vacuum of lattice gluodynamics behaves  as a dual superconductor:
the monopole currents are condensed and they are responsible for
confinement of color.

We have to note that  the approach discussed above is not unique and  there are several other approaches to  the confinement
problem in non-abelian gauge theories.
 We cannot discuss all these here, but
 mention  the description of confinement in terms
of $Z_2$ vortices \cite{Green}, the description of the QCD
vacuum in terms of the non-abelian dual lagrangians \cite{Baker},  and
the study of QCD by means of the  cumulant expansion \cite{Simonov}.

\section{Acknowledgments}

Authors are grateful to E.T.~Akhmedov, P.~van~Baal, F.V.~Gubarev,
T.L.~Ivanenko, Yu.A.~Simonov and T.~Suzuki for useful discussions.
M.N.Ch and M.I.P.  acknowledge the kind hospitality of the
Theoretical Department of the Kanazawa University. This work has been
supported by the JSPS Program on Japan -- FSU scientists
collaboration, and also by the Grants:  INTAS-94-0840, INTAS-94-2851,
INTAS-RFBR-95-0681, and Grant No.~96-02-17230a  of  the Russian
Foundation for Fundamental Sciences.

\section{Appendix~A: Differential Forms on the Lattice}
\setcounter{equation}{0}
\def\theequation{A.\arabic{equation}}

        Here we briefly summarize the main notions from the
theory of differential forms on the lattice. The calculus of differential
forms was developed for field theories on the lattice in
ref. \cite{deu}.
Since all lattice formulas have a direct continuum interpretation,
the standard formalism of differential geometry can have wide applications
in lattice theories. The advantage  of the calculus of differential forms
consists   in the general character of the obtained expressions. Most of the
transformations  depend  neither on the space--time dimension nor on the rank
of the fields.  With minor modifications,  the transformations are valid for
lattices of any form (triangular, hypercubic, random, etc).

        A differential form of rank $k$ on the lattice is a
skew-symmetrical function  $\phi_{k}$  defined on  $k$-dimensional
cells  $C_k$  of the lattice. The scalar lattice field is a 0--form. The
$U(1)$ gauge field is a 1--form. The exterior differential operator  {\it
d} is defined as follows:
\beq
(\dd \phi ) (C_{k+1}) =\sum_{\CK{k}' \in \partial\CK{k+1}}
        \phi(C_{k}')               \label{def-dd}
\eeq
Here $\partial C_{k}$ is the boundary of the $k$-cell $C_{k}$.
Thus,  the operator {\it d} increases the rank of the form by one. For
instance, in Wilson's $U(1)$ theory with the dynamic variables $U =
e^{i\theta}$, the action $\cos (\dd \theta)$ depends on the plaquette $\dd
\theta (C_2)$ constructed from the  links $\theta (C_1)$.

The dual lattice is defined as follows. The sites of the dual lattice are
placed at the centers of the D--dimensional cells of the original lattice.
The object dual to an oriented $k$-cell ($C_{k}$) is an oriented
$(D\!-\!k)$--cell ($\dual C_{k}$), which lies on the dual lattice and
has an intersection with $C_k$. The dual of the cubic lattice is also a
cubic lattice  obtained by shifting the lattice along each axis by
$1/2$ of the lattice spacing.
Every $k$-form on the lattice corresponds to a $(D\!-\!k)$--form on
the dual lattice:  $\dual\phi(\dual C_{k}) = \phi (C_k)$.  The
co-derivative is defined as follows:
\beq
         \delta = \dual\dd\dual.    \label{def-del}
\eeq
This operator decreases the rank of the form
\beq
(\delta\phi)(C_{k-1}) =
            \sum_{\mbox{\it all\ }\CK{k} :\CK{k-1}\in\partial\CK{k}}
                                                      \phi(C_{k}).
\eeq
The square of $\dd$ and $\delta$ is equal to zero:
\beqn
        \dd^2 & = & 0, \nonumber\\
        \delta^2 & = & 0.   \label{d2eq0}
\eeqn
The first equality is a consequence of the well-known geometrical fact:
``the boundary of the boundary is the empty set'';  the second equality
follows from the first one and  the definition of $\delta$
(\ref{def-del}).

        Now we discuss the lattice version of the Laplace operator, which is
defined as
\beq
\Delta = \delta\dd+\dd\delta.   \label{laplas}
\eeq
This operator acts on 0--forms (scalar fields) in the same way
as the usual finite-difference version of the continuous Laplacian. For
the forms of non-zero rank the relation  \eq{laplas} is the
generalization of the usual Laplacian. The obvious properties of $\delta$
are
\beqn
   \delta\Delta & = & \Delta\delta       \nonumber\\
   \dd\Delta & = & \Delta\dd             \label{propdel} \\
    1=\,\delta\Delta^{-1}\dd & + & \dd\Delta^{-1}\delta \nonumber
\eeqn
The last relation, called the Hodge identity, implies the  widely used
decomposition formula for an arbitrary $k$--form:
\beq
\phi=\delta\Delta^{-1}(\dd\phi)+\dd\Delta^{-1}(\delta\phi)+\phi_{h},
                                                             \label{Hodge}
\eeq
where $\phi_{h}$ is a harmonic form, $\delta\phi_{h} = \dd\phi_{h} = 0$.
The number of the harmonic forms depends on the topology of the space-time;
for instance, the number of the harmonic 1--forms is equal to 0,1 and D
respectively, for the space--time topology ${R}^{\DIM}$,
${R}^{\DIMO}\otimes {S}^1$ and $ {S}^{\DIM}$.

For two forms of the same rank the scalar product is introduced
in a natural way:
\beq
    (\phi,\psi) = \sum_{\mbox{\it all\ }\CK{k}}
    \phi(C_{k})\psi(C_{k})
\eeq
This scalar product agrees with the definitions of the $d$ and $\delta$
operators in the sense that the following formula of ``integration by
parts'' is valid:
\beq
(\dd\phi,\psi)=(\phi,\delta\psi).
\eeq
The norm of a $k$-form is defined as usual by
\beq
\| \phi \|^2 = \sum_{\CK{k}}
        \phi(C_k)^2  = (\phi,\phi) \label{norm}
\eeq
We illustrate the above definitions by four simple examples.

        {\bf Example 1.} Let us show  that the general action for the
compact gauge field is a periodic function of the angle $\bar{\theta}_P$
corresponding to the plaquette. By definition, $\bar{\theta}_P = \dd\theta + 2\pi k$,
where the integer valued 2--form $k$ is defined in such a way that $-\pi <
\bar{\theta}_P \leq \pi$. Under a gauge transformation we have $\theta'=\theta
+\dd\chi$ and $\bar{\theta}_P'=\dd\theta+{\dd}^2\chi+2\pi k'=\dd\theta+
2\pi k'$, where $k'$ is an integer  chosen in such a way that $-\pi <
\theta' \leq \pi$. Therefore,  the gauge invariance requires the periodicity
condition on the action:  $S(\bar{\theta}_P) = S(\bar{\theta}_P + 2\pi k)$.

  {\bf Example 2.} Note, that the definition of the norm
\eq{norm}  allows us to write in concise form the action of the
lattice theory, e.g.,  $S = \sum_{\CK{2}}
[\dd (A(C_1))]^2 = \| \dd A \|^2$ for noncompact electrodynamics.
We often use a similar notation for the Villain
action.

  {\bf Example 3.} Let us  show  that in standard compact lattice
electrodynamics there exists a conservation law. As we have mentioned
before,   the compact character of the fields implies the
existence of monopoles. The monopole charge inside an elementary
3-dimensional cube is defined by  $m = \frac {1}{4\pi}\dd\bar{\theta}_P =
\frac 12\dd k$, where $\bar{\theta}_P = \dd\theta + 2\pi k$, as in Example 1.
In the continuum limit,  the above definition corresponds to the Gauss
law:  $m = \frac {1}{4\pi} \oint_S {\vec H} \dd {\vec S}$, where $S$
is the surface of the elementary cube. If $j = \dual m$,  then
$\delta j = \dual \dd \dual \dual m = \frac 12\dual {\dd}^2 k = 0$.
Since $m$ is a rank 3 form, it follows that  $j$ is a rank 1 form and
the equation \beq \delta j = 0   \label{conserv} \eeq means that for
each site the incoming current $j$ is equal to the outgoing one.
Therefore,  the monopole current $j$ is conserved. It is easy to
prove that $j$ is gauge invariant. This result is well known for
4-dimensional lattice QED \cite{BMK}. Note that all transformations
considered above are valid for any space-time dimension $D$ and for the
field $\theta$ of any rank $k$; the current $j$ in this case is a
$(D\!-\!k\!-\!2)$--form. For the $XY$ model in $D = 3$ we have  $k = 0$ and the
conservation law (\ref{conserv}) for the 1-form $j$ implies that the
excitations, called vortices, form closed loops. In the 4-dimensional
$XY$ model the conserved quantity $j$  is represented by a 2-form,
which means that there exist excitations forming closed surfaces.
These objects are related to ``global'' cosmic strings, forming
closed surfaces \cite{PolWi90} in 4-dimensional space--time.

  {\bf Example 4.} To perform the BKT transformation (see Appendix
C) we have to solve the so-called cohomological equation
\beq
        \dd l=n,        \label{ddln}
\eeq
where $n$ is a given $k$--form and $l$ is a $(k-1)$-form to be found.
Using the Hodge decomposition, we can easily show
that $l(n)=\Delta^{-1}\delta n$ is a particular solution  which,  being added
to the general solution of the homogeneous equation, yields  the general
solution of \eq{ddln}: $l = \Delta^{-1}\delta n + \dd m$ ($m$ is an
arbitrary $(k-2)$-form).

If $n$ is an integer valued form, then the solution  $l$  of
\eq{ddln} can also be found in terms of integer numbers. We
explicitly construct such a solution for $k=2$. The generalization
for arbitrary values of $k$ is obvious. Let us assign the value $0$
to the links which belong to a given maximal tree on the lattice (a
maximal tree is a maximal set of links which do not contain closed
loops). To each link that does not belong to the tree we attach the value
equal to the value of the plaquette formed by the link and the tree
(there exists  one and only one such plaquette). It is easy to see
that we have thus constructed the required particular solution
$l(n)$.  The general solution has the form $l=l(n)+\dd m\ ,\ m \in
\Z$. For simplicity, we have neglected finite volume effects
(harmonic forms) in the above construction.

\section{Appendix~B: Duality transformation}
\setcounter{equation}{0}
\def\theequation{B.\arabic{equation}}

In this Appendix we perform the duality transformation for
$U(1)$ compact gauge theory with an arbitrary form of the action. Initially
the duality transformation was discussed by Kramers and
Wannier \cite{KrWa} for the 2--dimensional Ising model.

Let us consider the $U(1)$ gauge theory with an arbitrary periodic
action $S[\dd \theta]$, $S[\dots, X_P+2 \pi, \dots] = S[\dots, X_P,
\dots]$, where $\theta_l$ is the compact $U(1)$ link field and $P$
denotes any plaquette of the original lattice. We use the
formalism of differential forms  on the lattice (see Appendix~A).
We start from the partition function of the compact $U(1)$ theory:
\beqn
\cZ = \intpi \dD \theta \exp\Bigl\{- S[\dd \theta] \Bigr\}\,.
\label{initial}
\eeqn
The  Fourier expansion of the function $e^{- S(X)}$
 yields
\beqn
     \cZ = \const \intpi \dD \theta \nsum{k}{2}
     e^{-S^d (k)} e^{i (k,{\rm d} \theta) }\,, \label{sum}
\eeqn
where $k$ is an integer valued two-form, and the action $S^d$ is
\beqn
e^{-S^d (k)} = \const \intpi \dD X \exp\Bigl\{- S[X]
-i (k, X) \Bigr\}\,. \label{FT}
\eeqn
The integration over the field $\theta$ in eq.\eq{sum} gives the
constraint $\delta k = 0$,  which can be resolved by introducing the
new 3-form $n$: $k = \delta n$.   For simplicity, it is assumed here
that we are dealing with a lattice having trivial topology (e.g.
$\cR^4$). In the case of a lattice with a nontrivial topology,
arbitrary harmonic forms must be added to the r.h.s. of this
equation. Finally, changing the summation $\nsum{k}{2} \to
\ndsum{n}{1}$,  we get the dual representation of compact $U(1)$
gauge theory:
\beqn
 \cZ \propto \cZ^d = \ndsum{n}{1} e^{-S^d (\dual {\rm d} n)}\,.
 \label{dsum}
\eeqn

Consider now compact $U(1)$ gauge theory with the action in the
Villain form eq.\eq{Villain}. Now the integral is Gaussian and the
dual action  is $S^d (\dual\dd n)={||\dual\dd n||}^2/4\beta$.

\section{Appendix~C: The BKT transformation}
\setcounter{equation}{0}
\def\theequation{C.\arabic{equation}}

In this Appendix we show how to perform the BKT transformation for
compact $U(1)$ gauge theory with an  arbitrary form for the
action~\eq{initial}. Let us insert the unity
\beqn 1 = \intinf \dD G \, \delta(G - k) \nonumber
\eeqn
(here $G$ is a real-valued
two-form) into the sum (\ref{sum})
\beqn \cZ = \intpi \dD \theta
\exp\Bigl\{- S_P(\dd \theta) \Bigr\} = \const\!\! \intpi \dD \theta
\intinf \dD G\!\! \nsum{k}{2}\!\! \delta(G-k) e^{-S^d(G)}  e^{i (G, {\rm d}
\theta)}.
\eeqn
Using the Poisson summation formula
\beqn
\nsum{k}{2} \delta(G-k) = \nsum{k}{2} e^{2 \pi i (G, k)}\,,
\eeqn
we get
\beqn \cZ = \const \intpi \dD \theta  \intinf \dD G
\nsum{k}{2} e^{-S^d(G)}\, e^{i ({\rm d} \theta + 2 \pi k, G)}\,.
\label{poisson}
\eeqn

Now we perform the BKT transformation with respect to the
integer valued 2-form~$k$:
\beqn
 k = m[j] + \dd q\,,\quad \dd m[j] = j\,,\quad \dd j =0\,,
 \label{bkt0}
\eeqn
where $q$ and $j$ are one- and three-forms,  respectively. First,
we change the summation variable, $\displaystyle{\nsum{k}{2} =
\sum_{\stackrel{\scriptstyle j (\CK{3}) \in \Z} {{\rm d} j=0}}
\nsum{q}{1}}$. Using the Hodge--de--Rahm decomposition,  we adsorb the
d--closed part of the 2--form $k$ into the compact variable $\theta$:
\beqn
 \dd \theta + 2 \pi k = \dd \theta_{n.c.}
 + 2 \pi \delta {\Delta}^{-1} j\,,
 \quad \theta_{n.c.} = \theta + 2 \pi
 {\Delta}^{-1} \delta m[j] + 2 \pi q\,. \label{bkt}
\eeqn

Substituting eq.(\ref{bkt}) in eq.(\ref{poisson}) and integrating over
the noncompact field $\theta_{n.c.}$,  we get the following
representation of the partition function:
\beqn
        \cZ = \const \intinf \dD G \nddsum{j}{1}
        e^{-S^d(G)}\, \exp\Bigl\{2 \pi i (G, \delta
        {\Delta}^{-1} j) \}\, \delta(\delta G)\,. \label{step3}
\eeqn
The constraint $\delta G = 0$ can be solved by $G = \delta
H$, where $H$ is a real valued 3--form.  Substituting this solution
into eq.(\ref{step3}) we obtain  the final
expression for the BKT--transformed action on the dual lattice:
\beqn
        \cZ = \const \nddsum{j}{1} e^{- S_{mon}(\dual j)}\,.
        \label{GenFin:1}
\eeqn
where
\beqn
 S_{mon}(\dual j) = - \ln \left(\,\, \intinf \dD H
 e^{-S^d (\delta H)}\,
 \exp\Bigl\{2 \pi i (\dual H,\dual j) \Bigr\} \right)\,.
 \label{GenFin:2}
\eeqn
We have used the relation $\dd \delta {\Delta}^{-1} j \equiv j,\ \forall
j: \dd j = 0$.  Therefore,  for the general $U(1)$ action $S[\dd
\theta]$ the monopole action \eq{GenFin:2} is nonlocal and it is
expressed through the integral over the  entire  lattice ($\intinf \dD H$).

As an example we consider the Villain form of the $U(1)$ action
\eq{Villain}. Repeating all the steps we get the
following monopole action:
\beqn
S_{mon}(\dual j) = 4 \pi^2 \beta (\dual j, \Delta^{-1} \dual j)\,.
\eeqn

\end{document}